\documentclass[a4paper,10pt]{article}
\usepackage{lmodern}
\usepackage{helvet}
\usepackage{graphicx}
\usepackage{float}
\usepackage[font=footnotesize]{caption}
\usepackage{subcaption}
\usepackage{amsmath, amssymb}
\usepackage{mathtools}
\usepackage[per-mode=symbol]{siunitx}
\usepackage{booktabs}
\usepackage{changepage}          
\usepackage{geometry}
\usepackage{setspace}
\usepackage{titlesec}
\titlespacing{\section}{0pt}{10pt}{0pt}
\titlespacing{\subsection}{0pt}{5pt}{0pt}
\titlespacing*{\paragraph}{0pt}{2pt}{5pt}
\usepackage{fancyhdr}
\usepackage[colorlinks, linkcolor=blue, urlcolor=blue, citecolor=blue, bookmarksnumbered]{hyperref}
\usepackage[style=phys, biblabel=brackets]{biblatex}
\begin{document}
\thispagestyle{plain}

\begin{center}
    \Large\textbf{Surrogate-assisted three-dimensional Gross–Pitaevskii characterization of Bose–Einstein condensate transport on an atom chip}\\[2ex]
    \normalsize Naoki Shibuya\\
    Independent Researcher, Tokyo, Japan
\end{center}

\begin{center}
    \textbf{\Large Abstract}
\end{center}

\begin{adjustwidth}{2cm}{2cm}
    We develop a surrogate-assisted workflow for millimeter-scale transport of a $^{87}$Rb Bose--Einstein condensate (BEC) on a multi-wire atom chip. Gradient-based inverse optimization generates 42 current schedules spanning seven regularization strengths and six transport durations. A low-cost Gaussian phase-space model ranks these schedules by endpoint overlap at three atom numbers. A Thomas--Fermi scaling (Ermakov) model estimates cloud dimensions for case-specific moving-grid construction. We then evaluate selected schedules in detail with the three-dimensional Gross--Pitaevskii equation (GPE). At $N=10^3$ and $T=\SI{0.5}{\second}$, the surrogate predicts a larger axial center-of-mass (COM) excursion than at longer durations. The GPE confirms this prediction and reveals sloshing that persists through delivery, with an endpoint fidelity $F_{3D}=0.9244$. For all five longer transports, the surrogate predicts near-unit endpoint overlap for the top-ranked schedules, and GPE propagation using these schedules gives $F_{3D}>0.9997$. At $N=10^4$ and $T=\SI{2.0}{\second}$, the surrogate-selected schedule similarly gives $F_{3D}=0.999932$. These results demonstrate that the surrogate estimates provide a useful low-cost basis for candidate screening, while GPE propagation resolves the transport dynamics and the delivered condensate state.
\end{adjustwidth}

\section{Introduction}

Atom chips use microfabricated current-carrying wires and external bias fields to confine and transport ultracold gases near surfaces~\cite{reichel2011trapping}. In BEC microscopy, for example, the condensate is transported from the initial trapping region to a separate probing region~\cite{feketeQuantumGasEnabledDirect2024}. This work adopts the magnetic-conveyor scheme of Long et al.~\cite{long_2005_LongDistanceMagnetic}, in which modulated currents translate a chain of Ioffe--Pritchard-type wells. We use regularized inverse optimization to determine the wire-current schedule for moving the trap minimum along a prescribed straight path~\cite{shibuya2025arxiv}. This prioritizes explicit control over the path, unlike shortcut-to-adiabaticity trajectory design aimed at suppressing residual excitation~\cite{corgierFastManipulationBose2018a}.

However, our previous study~\cite{shibuya2025arxiv} assessed the schedules only through trap-geometry metrics and a velocity-based adiabaticity criterion. These measures do not reveal how the cloud evolves during transport or determine how well its final wavefunction overlaps with the destination ground state. GPE propagation~\cite{dalfovo1999theory} can resolve acceleration-driven sloshing and mean-field dynamics, but applying it to every combination of regularization strength, transport duration, and atom number would be computationally expensive. We therefore use surrogate models to explore the candidate space and identify informative cases for GPE characterization. Their estimates also help automate moving-grid construction and time-step selection, thereby avoiding case-by-case manual tuning.

Overall, this surrogate-assisted workflow combines efficient exploration of wire-current designs with three-dimensional GPE simulation of BEC transport.

\section{Candidate Schedule Generation}\label{sec:control-schedule}

The atom-chip geometry, magnetic-field model, and inverse-optimization procedure follow our prior work~\cite{shibuya2025arxiv}. The details required for the present study are summarized below.

\subsection{Atom-Chip Geometry and Trap Potential}
\begin{figure}[!htbp]
    \centering
    \includegraphics[width=0.95\textwidth]{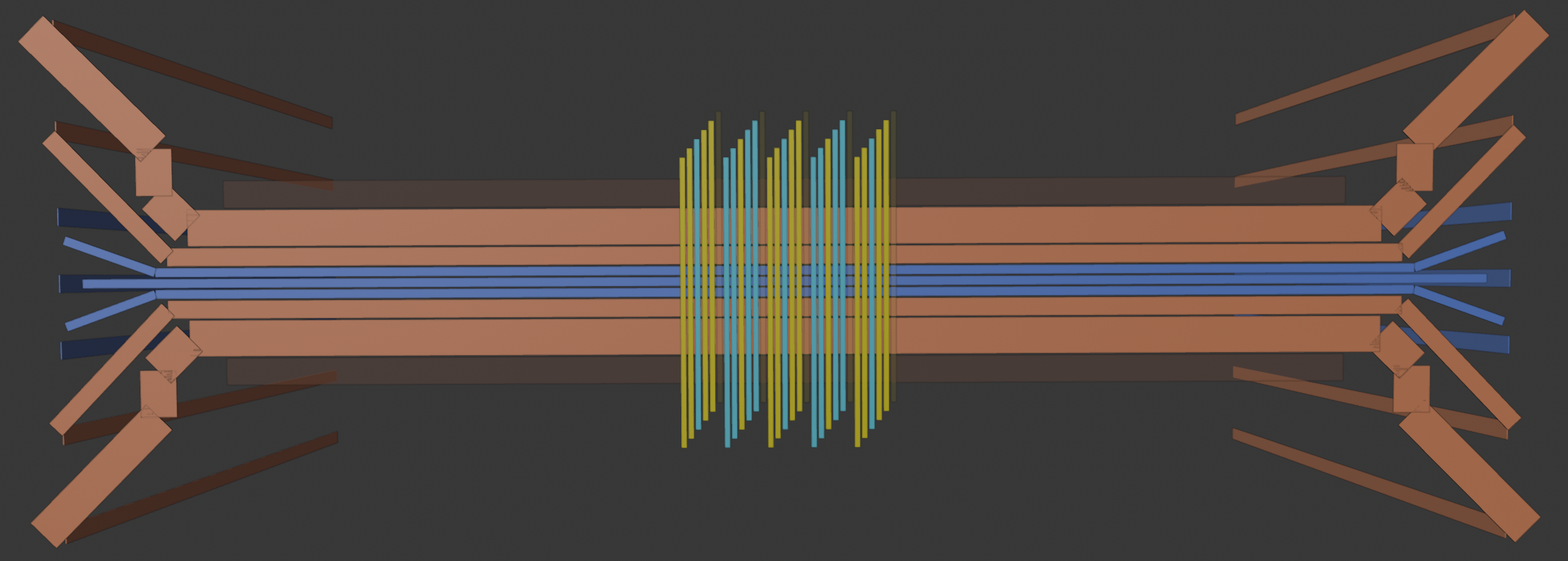}
    \caption{Three-dimensional atom-chip layout. Shifting wires are gold and guiding wires copper; segments carrying reverse current are tinted cyan or blue. The condensate is transported by one conveyor period along the guiding-wire direction.}
    \label{fig:chip-layout}
\end{figure}

The magnetic conveyor uses the multi-wire atom-chip geometry shown in Fig.~\ref{fig:chip-layout}. We take $\hat{x}$ along the transport direction, $\hat{z}$ normal to the chip surface and pointing into the trapping region, and $\hat{y}$ completing the right-handed triad.
The chip uses 15 independently controlled wire currents. Nine guiding wires running along $\hat{x}$ provide transverse quadrupole confinement, while six shifting wires crossing the transport axis translate the trap minimum along $\hat{x}$. The guiding-wire currents remain fixed, while only the shifting-wire currents are varied, with each constrained by $|I_i|\leq\SI{3.5}{\ampere}$.

We compute the magnetic field by applying the Biot–Savart law to conducting wires modeled as finite rectangular prisms, following the approach in~\cite{barrettApparatusProductionBoseEinstein2017}. For $^{87}$Rb in the low-field-seeking $|F{=}2,m_F{=}2\rangle$ sublevel, the magnetic-plus-gravity potential is
\begin{equation}
    U(\mathbf r, t)
    =
    g_F \, m_F \, \mu_B \, \left|\mathbf B(\mathbf r, t)\right| \,
    -
    \, mgz,
\end{equation}
where $g_F=1/2$, $\mu_B$ is the Bohr magneton, $m=\SI{1.44e-25}{\kilo\gram}$ is the atomic mass, and $g$ is the gravitational acceleration. The coordinate $z$ is measured downward from the chip surface, so the gravitational potential decreases with increasing $z$, giving the term $-mgz$.

The harmonic approximation of the trap potential about its minimum $\mathbf r_{\min}(t)$ is
\begin{equation}
    U(\mathbf r,t)
    \approx
    U(\mathbf r_{\min}(t),t)
    +
    \frac{1}{2}
    \left[
        \mathbf r-\mathbf r_{\min}(t)
    \right]^\top
    H(t)
    \left[
        \mathbf r-\mathbf r_{\min}(t)
    \right],
    \label{eq:harmonic-potential}
\end{equation}
where $H(t)$ is the Hessian matrix of second derivatives of the potential evaluated at the trap minimum,
\begin{equation}
    H_{ij}(t)
    =
    \left.
        \frac{\partial^2 U(\mathbf r,t)}
            {\partial r_i\,\partial r_j}
    \right|_{\mathbf r=\mathbf r_{\min}(t)},
    \qquad
    i,j\in\{x,y,z\}.
    \label{eq:trap-hessian}
\end{equation}
The eigenvectors $\hat{\mathbf e}_k(t)$ and eigenvalues $m\omega_k^2(t)$ define the trap principal axes and angular frequencies.
\begin{equation}
    H(t)\,\hat{\mathbf e}_k(t)
    =
    m\,\omega_k^2(t)\,\hat{\mathbf e}_k(t),
    \qquad
    k\in\{1,2,3\}.
\end{equation}
We order the axes by increasing frequency, so that $k=1$ denotes the longitudinal direction, and write the corresponding frequencies in hertz as $f_k=\omega_k/(2\pi)$.

\begin{figure}[!htbp]
    \centering
    \begin{subfigure}[t]{0.48\textwidth}
        \centering
        \includegraphics[width=0.95\linewidth]{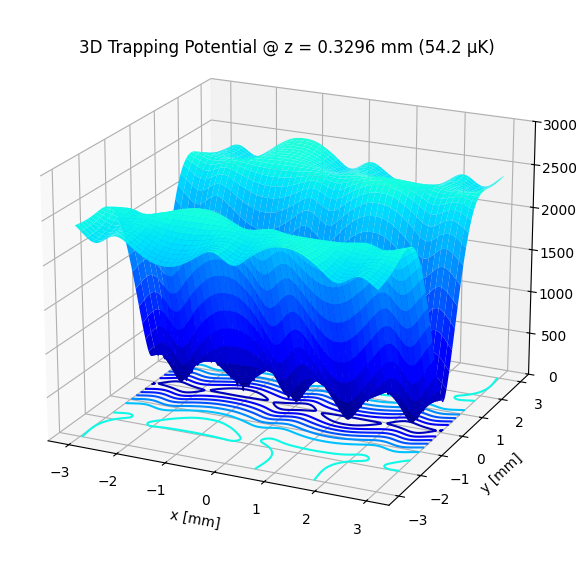}
        \caption{Potential surface.}
    \end{subfigure}
    \hfill
    \begin{subfigure}[t]{0.48\textwidth}
        \centering
        \includegraphics[width=0.95\linewidth]{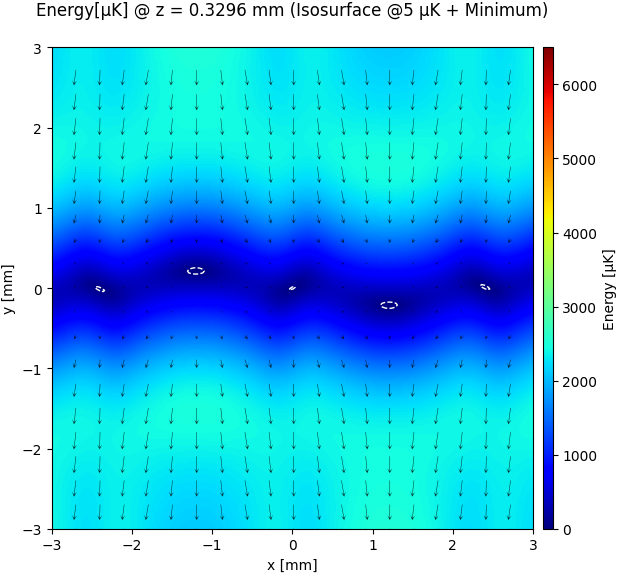}
        \caption{Contour map.}
    \end{subfigure}
    \caption{The central well is selected as the initial trap, and its minimum is followed over one conveyor period.}
    \label{fig:conveyor-potential}
\end{figure}

The wire configuration produces a periodic chain of Ioffe--Pritchard-type wells along the transport direction, each with a finite magnetic-field minimum that suppresses Majorana spin flips (Fig.~\ref{fig:conveyor-potential}). 

We select the central well as the initial trap and track its local minimum over one conveyor period spanning six shifting wires. The initial trap minimum lies at $z_{\min}=\SI{0.3296}{\milli\meter}$. The principal-axis trap frequencies are $(f_1,f_2,f_3)\simeq(110.1,343.3,362.2)\,\si{\hertz}$, showing the stronger transverse confinement.

\subsection{Schedule Generation}

All schedules begin from the initial trap and translate the trap minimum by one conveyor period, $\Delta x=\SI{2.4}{\milli\meter}$, along $\hat{x}$ over a total duration $T$. Each schedule consists of shifting-wire current vectors $\{\mathbf I_j\}$ specified at $n_{\rm ctrl}+1$ uniformly spaced control nodes,
\begin{equation}
    t_j=j\Delta t_{\rm ctrl},
    \qquad
    j=0,\ldots,n_{\rm ctrl},
    \qquad
    T=n_{\rm ctrl}\Delta t_{\rm ctrl}.
\end{equation}
We set the control interval to $\Delta t_{\rm ctrl}=\SI{5e-4}{\second}$ (a control cadence of \SI{2}{\kilo\hertz}) as a practical choice balancing temporal resolution against the number of inverse-optimization steps. At control node $j$, the prescribed axial path follows the quintic smoothstep $10s_j^3-15s_j^4+6s_j^5$, where $s_j=t_j/T$. Its first and second derivatives vanish at both endpoints.

Each current vector $\mathbf I$ determines the external potential $U(\mathbf r)$ and its trap minimum $\mathbf r_{\min}$. At each optimization step, the desired displacement $\delta\mathbf r$ of the trap minimum is mapped to a current update $\delta\mathbf I$,
\begin{equation}
    \delta\mathbf I
    =
    (J^\top J+\alpha\mathbb I)^{-1}J^\top\delta\mathbf r,
    \qquad
    \alpha
    =
    \lambda\bigl(1+\kappa(J)\bigr),
    \qquad
    \kappa(J)
    =
    \frac{\sigma_{\max}(J)}{\sigma_{\min}(J)},
    \label{eq:tikhonov}
\end{equation}
where $\mathbb I$ is the identity matrix, $J=d\mathbf r_{\min}/d\mathbf I$ is the trap-minimum Jacobian, and $\kappa(J)$ is its condition number, with $\sigma_{\max}(J)$ and $\sigma_{\min}(J)$ denoting its largest and smallest singular values. The regularization parameter $\lambda$ sets the balance between displacement accuracy and current-update magnitude.

\begin{figure}[!htbp]
    \centering
    \includegraphics[width=0.9\textwidth]{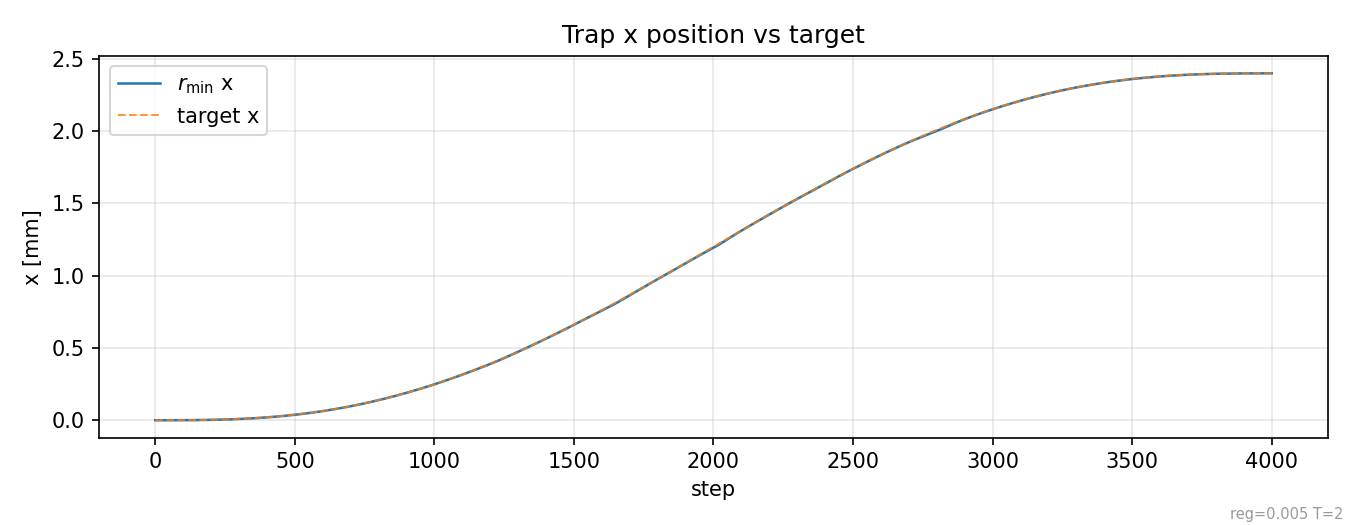}
    \caption{Representative optimized trajectory for $\lambda=5.0\times10^{-3}$ and $T=\SI{2.0}{\second}$. The optimized trap-minimum position $r_{\min,x}(t_j)$ closely follows the prescribed quintic target over 4000 control intervals.}
    \label{fig:position-x}
\end{figure}

\begin{figure}[!htbp]
    \centering
    \includegraphics[width=1.0\textwidth]{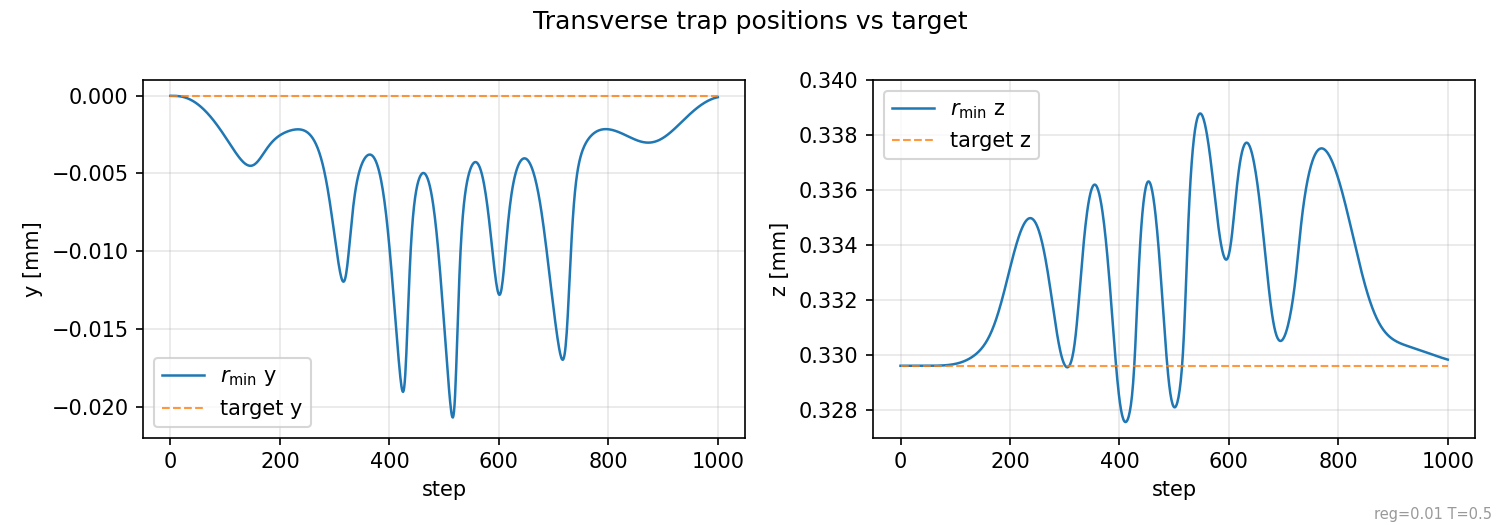}
    \caption{Transverse trap-minimum trajectories for $\lambda=10^{-2}$ and $T=\SI{0.5}{\second}$.}
    \label{fig:transverse-trajectories1}
\end{figure}

\begin{figure}[!htbp]
    \centering
    \includegraphics[width=1.0\textwidth]{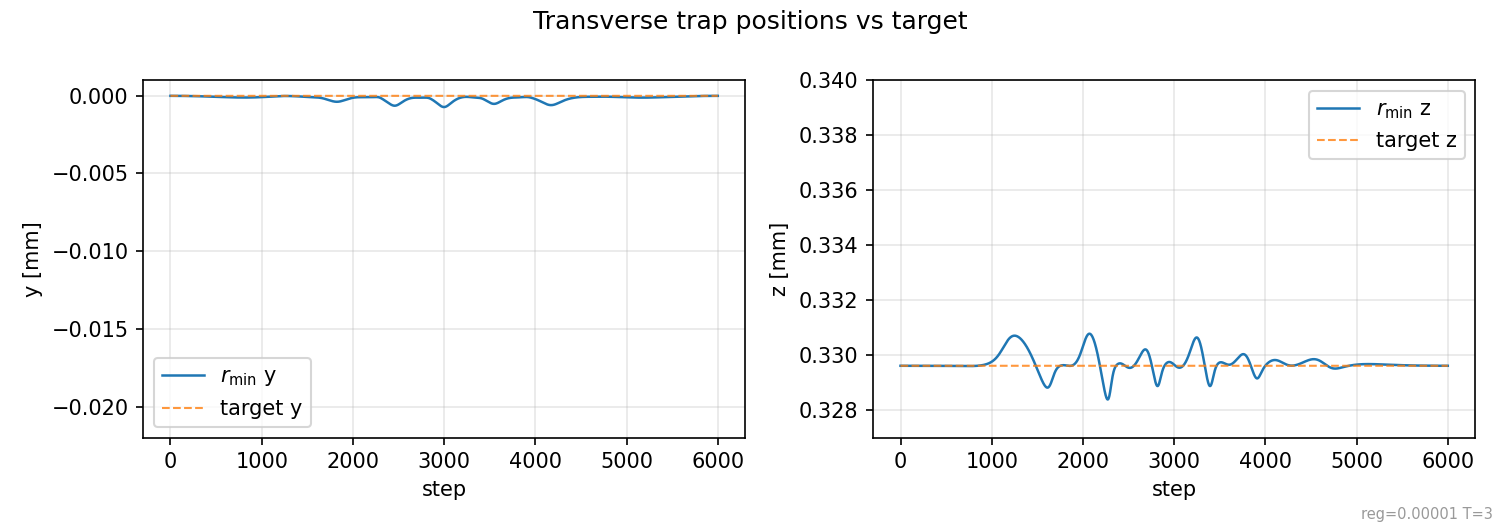}
    \caption{Transverse trap-minimum trajectories for $\lambda=10^{-5}$ and $T=\SI{3.0}{\second}$.}
    \label{fig:transverse-trajectories2}
\end{figure}

The inverse solver generates one candidate schedule for each $(\lambda,T)$ pair. We consider the following regularization values
$\Lambda =
    \{
        0.001,\,
        0.005,\,
        0.01,\,
        0.05,\,
        0.1,\,
        0.5,\,
        1.0
    \} \times 10^{-2}\ \si{\milli\meter\squared\per\ampere\squared}
$
and transport durations 
$\mathcal T
    =
    \{
        0.5,\,
        1.0,\,
        1.5,\,
        2.0,\,
        2.5,\,
        3.0
    \}\,\si{\second}.
$
Fig.~\ref{fig:position-x} shows the resulting trajectory for a representative candidate.
Although the prescribed path is purely axial, the optimized currents also produce transverse motion of the trap minimum. Fig.~\ref{fig:transverse-trajectories1} and Fig.~\ref{fig:transverse-trajectories2} compare two contrasting $(\lambda,T)$ pairs and show substantial differences in both the magnitude and temporal variation of this motion.

We therefore account for displacement and velocity of the condensate center of mass (COM) by treating the condensate dynamics in three dimensions.

\section{GPE Model and Numerical Methods}\label{sec:gpe-model}

\subsection{Gross--Pitaevskii Model}\label{sec:gpe}

We model the condensate as a macroscopic wavefunction $\psi$ evolving under the time-dependent GPE
\begin{equation}
    i\hbar\, \frac{\partial \psi(\mathbf{r}, t)}{\partial t}
    =
    \left[
    -\frac{\hbar^2}{2m} \nabla^2
    +
    U(\mathbf{r}, t)
    +
    g_{\rm eff} |\psi|^2
    \right]
    \psi,
    \label{eq:gpe}
\end{equation}
where $U(\mathbf r, t)$ is the external potential and $g_{\rm eff}|\psi|^2$ is the mean-field interaction (Appendix~\ref{appendix:gpe-derivation}). We normalize the wavefunction such that $\int |\psi|^2\,d^3\mathbf r=1$. The atom density is then $N|\psi|^2$, where $N$ is the atom number. For large $N$, the coefficient $g_{\rm eff}$ is approximated as
\begin{equation}
    g_{\rm eff}
    \simeq
    \frac{4 \pi \hbar^2 a_s N}{m},
    \label{eq:g-effective}
\end{equation}
where $a_s$ is the $s$-wave scattering length. For $^{87}$Rb, $a_s=\SI{5.2e-9}{\meter}>0$, so the interaction is repulsive.

To characterize the cloud dynamics and fidelity, we compute the following observables.

\paragraph{Center of Mass.}
The condensate COM displacement relative to the trap minimum $\mathbf r_{\min}(t)$ is
\begin{equation}
    \mathbf r_{\rm com}(t)
    =
    \bar{\mathbf r}(t)-\mathbf r_{\min}(t),
    \qquad
    \bar{\mathbf r}(t)
    =
    \int \mathbf r\,|\psi(\mathbf r, t)|^2\,d^3\mathbf r.
    \label{eq:gpe-com}
\end{equation}
Oscillations of $\mathbf r_{\rm com}(t)$ characterize sloshing.

\paragraph{Cloud Geometry.}
We characterize the condensate density using its covariance matrix
\begin{equation}
    \Sigma(t)
    =
    \int
    \bigl(\mathbf r-\bar{\mathbf r}(t)\bigr)
    \bigl(\mathbf r-\bar{\mathbf r}(t)\bigr)^\top
    \ |\psi(\mathbf r, t)|^2
    \,d^3\mathbf r .
    \label{eq:gpe-covariance}
\end{equation}
Diagonalizing $\Sigma(t)$ gives the cloud principal axes $\hat{\mathbf c}_j(t)$ and their variances $\lambda_j(t)$,
\begin{equation}
    \Sigma(t)\,\hat{\mathbf c}_j(t)
    =
    \lambda_j(t)\,\hat{\mathbf c}_j(t),
    \qquad
    \left\lVert\hat{\mathbf c}_j(t)\right\rVert=1,
    \qquad
    j\in\{1,2,3\} .
\end{equation}

\paragraph{Cloud Orientation.}
We choose the permutation $\pi$ that gives the best match between the cloud and trap principal axes, so $\hat{\mathbf c}_{\pi(k)}(t)$ is paired with $\hat{\mathbf e}_k(t)$. Their unsigned misalignment is
\begin{equation}
    \theta_k(t)
    =
    \cos^{-1}\!\left(
        \left|
            \hat{\mathbf c}_{\pi(k)}(t)\cdot
            \hat{\mathbf e}_k(t)
        \right|
    \right),
    \label{eq:gpe-misalignment}
\end{equation}
and the cloud principal rms width paired with trap axis $k$ is
\begin{equation}
    \sigma_k(t)=\sqrt{\lambda_{\pi(k)}(t)} .
    \label{eq:sigma}
\end{equation}
For the Ioffe--Pritchard-type trap used here, the cloud's nearly circular transverse cross-section makes its transverse principal axes poorly conditioned, so we interpret only the axial orientation misalignment $\theta_1(t)$. We use the normalized axial width $\sigma_1(t)/\sigma_1(0)$ to characterize axial breathing during transport.

\paragraph{Endpoint Fidelity.}
The endpoint fidelity between the final state $\psi(\mathbf r,T)$ and the target ground state $\psi_{\rm target}(\mathbf r)$ is
\begin{equation}
    F_{3D}
    =
    \left|
        \int
        \psi_{\rm target}^*(\mathbf r)\,
        \psi(\mathbf r,T)
        \,d^3\mathbf r
    \right|^2 .
    \label{eq:gpe-endpoint-fidelity}
\end{equation}
$F_{3D}=1$ only when the two states are identical except for a global phase.

\subsection{Numerical Propagation}\label{sec:numerical-implementation}

We solve Eq.~\ref{eq:gpe} using a split-step Fourier method on a Cartesian grid translated with the trap minimum.

\paragraph{Currents Between Control Nodes.}
The optimized schedule specifies current vectors $\mathbf I_j$ at control times separated by $\Delta t_{\rm ctrl}$. GPE propagation uses finer time steps to resolve the condensate dynamics, so the applied current must be defined between these nodes. We hold each $\mathbf I_j$ constant over the interval centered on $t_j$, with switches at the midpoint $t_{j+\frac{1}{2}}=(t_j+t_{j+1})/2$. The applied current vector is then
\begin{equation}
    \mathbf I(t)
    =
    \begin{cases}
        \mathbf I_0,
        & 0\leq t<t_{\frac{1}{2}},\\[1ex]
        \mathbf I_j,
        & t_{j-\frac{1}{2}}\leq t<t_{j+\frac{1}{2}},
        \qquad j=1,\ldots,n_{\rm ctrl}-1,\\[1ex]
        \mathbf I_{n_{\rm ctrl}},
        & t_{n_{\rm ctrl}-\frac{1}{2}}\leq t\leq T .
    \end{cases}
    \label{eq:node-centered-current}
\end{equation}
Thus, the initial and final current vectors are each applied for half an interval, while every interior vector is applied for a complete interval centered on its control-node time. We choose the number of substeps $n_{\rm sub}\in2\mathbb N$ to place every midpoint current switch on a propagation-step boundary, so the time step is
\begin{equation}
    \Delta t
    =
    \frac{\Delta t_{\rm ctrl}}{n_{\rm sub}} .
    \label{eq:propagation-time-step}
\end{equation}

\paragraph{Split-Step Propagation.}\label{sec:strang-factorization}
We advance each propagation step using the symmetric Strang factorization
\begin{equation}
    \begin{aligned}
        \psi(\mathbf r, t+\Delta t)
        &=
        \exp\!\left(-i\frac{\hat T}{\hbar}\frac{\Delta t}{2}\right)
        \exp\!\left(-i\frac{\hat V}{\hbar}\Delta t\right)
        \exp\!\left(-i\frac{\hat T}{\hbar}\frac{\Delta t}{2}\right)
        \psi(\mathbf r, t)
        +
        \mathcal O(\Delta t^3),\\[1ex]
        &\text{where }\quad
        \hat T = -\frac{\hbar^2}{2m}\nabla^2,
        \quad
        \hat V=U(\mathbf r, t)+g_{\rm eff}|\psi|^2.
    \end{aligned}
    \label{eq:strang-step}
\end{equation}
The kinetic half-steps are applied pointwise on the wavevector grid using Fourier transforms. The potential phase is applied pointwise on the spatial grid using $g_{\rm eff}|\psi|^2$ evaluated at the beginning of the stage. The method has local error $\mathcal O(\Delta t^3)$ and global error $\mathcal O(\Delta t^2)$ (Appendix~\ref{appendix:operator-splitting}).

\paragraph{Spatial Grid and Integration.}
We discretize the wavefunction on a Cartesian grid. The Fourier transforms treat the grid values as periodic along each axis. The grid spacing is
\begin{equation}
    \Delta r_i
    =
    \frac{L_i}{N_i},
    \qquad
    i\in\{x,y,z\},
\end{equation}
where $L_i$ is the grid length along axis $i$ and $N_i$ is the cell count. Defining the volume element $\Delta V=\Delta r_x\,\Delta r_y\,\Delta r_z$ and grid index $\mathbf n=(n_x,n_y,n_z)$, we approximate the spatial integrals by grid sums
\begin{equation}
    \begin{aligned}
        \int f(\mathbf r)|\psi(\mathbf r, t)|^2\,d^3\mathbf r
        &\approx
        \sum_{\mathbf n}
        f_{\mathbf n}|\psi_{\mathbf n}(t)|^2\,\Delta V
        && (\text{for Eqs.}~\ref{eq:gpe-com}\text{ and}~\ref{eq:gpe-covariance}),
        \\
        \int
        \psi_{\rm target}^{*}(\mathbf r)
        \psi(\mathbf r,T)
        \,d^3\mathbf r
        &\approx
        \sum_{\mathbf n}
        \psi_{{\rm target},\mathbf n}^{*}
        \psi_{\mathbf n}(T)
        \,\Delta V
        && (\text{for Eq.~}\ref{eq:gpe-endpoint-fidelity}).
    \end{aligned}
\end{equation}

\paragraph{Wavevector Grid and Translation.}
The discrete Fourier transform expands the sampled wavefunction into plane waves $e^{i\mathbf k\cdot\mathbf r}$. The wavevector components are separated by
\begin{equation}
    \Delta k_i
    =
    \frac{2\pi}{L_i}.
\end{equation}
The three-dimensional wavevector grid contains every combination
\begin{equation}
    \mathbf k_{\boldsymbol\ell}
    =
    \left(
        \ell_x\Delta k_x,
        \ell_y\Delta k_y,
        \ell_z\Delta k_z
    \right),
    \qquad
    \ell_i
    \in
    \left\{
        -\left\lfloor\frac{N_i}{2}\right\rfloor,
        \ldots,
        \left\lceil\frac{N_i}{2}\right\rceil-1
    \right\},
    \qquad
    i\in\{x,y,z\}.
\end{equation}
We use Google's JAX for the discrete Fourier transforms. For example, its FFT array order for $N_i=8$ is
\begin{equation}
    \ell_i=(0,1,2,3,-4,-3,-2,-1).
\end{equation}
Because the transport distance is much larger than the condensate size, we shift the numerical grid origin to follow the trap minimum. At the midpoint switch from $\mathbf I_{j-1}$ to $\mathbf I_j$, the origin shifts by
\begin{equation}
    \Delta\mathbf r_{\min,j}
    =
    \mathbf r_{\min}(t_j)
    -
    \mathbf r_{\min}(t_{j-1}).
\end{equation}
The origin shift requires evaluating the wavefunction at the shifted spatial coordinates,
\begin{equation}
    \psi_{\rm new}(\mathbf r)
    =
    \psi_{\rm old}
    \left(
        \mathbf r+\Delta\mathbf r_{\min,j}
    \right).
\end{equation}
We calculate these shifted values in the wavevector representation. Let $\widetilde{\psi}(\mathbf k)$ denote the Fourier transform of $\psi(\mathbf r)$. We apply the shift by multiplying each Fourier component,
\begin{equation}
    \widetilde{\psi}_{\rm new}(\mathbf k)
    =
    \exp\!\left(
        i\mathbf k\cdot\Delta\mathbf r_{\min,j}
    \right)
    \widetilde{\psi}_{\rm old}(\mathbf k),
\end{equation}
and then apply the inverse transform to return the shifted wavefunction to the spatial representation.

\subsection{Imaginary-Time Relaxation}\label{sec:imaginary_time}

The initial and target ground states are obtained by iterative imaginary-time relaxation with discrete normalization~\cite{bao2004groundstate}, while the external potential is fixed at $t=0$ and $t=T$, respectively.

\paragraph{Initialization.}
For each relaxation, the Gaussian trial profile uses the Thomas--Fermi rms widths (Appendix~\ref{appendix:tf-equilibrium}) as initial scale parameters,
\begin{equation}
    \sigma_{{\rm seed},k}
    =
    \sigma_{{\rm TF},k}
    =
    \frac{R_{{\rm TF},k}}{\sqrt7},
    \qquad
    k\in\{1,2,3\}.
\end{equation}
Because the GPE has the $g_{\rm eff}$ term that includes the atom number under the convention $\int|\psi|^2\,d^3\mathbf r=1$, the trial profile is normalized before the first relaxation step so that the interaction term $g_{\rm eff}|\psi|^2$ has the intended scale from the start.

\paragraph{Split-Step Update.}
Each relaxation step advances the imaginary time from $\tau$ to $\tau+\Delta\tau$ and produces an unnormalized state,
\begin{equation}
    \psi'(\mathbf r, \tau+\Delta\tau)
    =
    \exp\!\left(-\frac{\hat T}{\hbar}\frac{\Delta\tau}{2}\right)
    \exp\!\left(-\frac{\hat V}{\hbar}\Delta\tau\right)
    \exp\!\left(-\frac{\hat T}{\hbar}\frac{\Delta\tau}{2}\right)
    \psi(\mathbf r, \tau),
\end{equation}
where $\hat V = U(\mathbf r,t) + g_{\rm eff}|\psi|^2$. The nonlinear contribution $g_{\rm eff}|\psi|^2$ is evaluated at the beginning of the potential stage and held fixed during that update. After the second kinetic half-step, the wavefunction is normalized as
\begin{equation}
    \psi(\mathbf r, \tau+\Delta\tau)
    =
    \frac{\psi'(\mathbf r, \tau+\Delta\tau)}
    {
        \sqrt{
            \int|\psi'|^2\,d^3\mathbf r
        }
    }.
\end{equation}
We repeat this relaxation step for a prescribed number of iterations $n_{\rm imag}$. The imaginary-time relaxation is detailed in Appendix~\ref{appendix:imaginary-time}.

\section{Surrogate-Assisted Screening}\label{sec:surrogate-models}

\subsection{Endpoint-Fidelity Estimate}\label{sec:gaussian-phase-space}

We estimate endpoint fidelity by approximating the delivered cloud and the target ground state as Gaussian profiles of the same size and shape, differing only in residual displacement and velocity. We calculate these quantities by propagating the harmonic COM equation (Appendix~\ref{appendix:gaussian-propagation}). At control node $j$, we combine the cloud centroid $\bar{\mathbf r}(t_j)$ and mean momentum $\bar{\mathbf p}(t_j)$ into the augmented vector
\begin{equation}
    \mathbf y_j
    =
    \begin{pmatrix}
        \bar{\mathbf r}(t_j)\\
        \bar{\mathbf p}(t_j)\\
        1
    \end{pmatrix},
    \qquad
    \mathbf y_0
    =
    \begin{pmatrix}
        \mathbf r_{\min}(0)\\
        \mathbf 0\\
        1
    \end{pmatrix},
\end{equation}
and define
\begin{equation}
    \mathcal A_j
    =
    \begin{pmatrix}
        0 & \mathbb I/m & 0\\
        -H_j & 0 & H_j\mathbf r_{\min,j}\\
        0 & 0 & 0
    \end{pmatrix},
    \qquad
    H_j=H(t_j),
    \qquad
    \mathbf r_{\min,j}=\mathbf r_{\min}(t_j).
\end{equation}
Under the node-centered current schedule (Eq.~\ref{eq:node-centered-current}), the vectors at adjacent control nodes satisfy
\begin{equation}
    \mathbf y_{j+1}
    =
    \exp\!\left(
        \mathcal A_{j+1}\frac{\Delta t_{\rm ctrl}}{2}
    \right)
    \exp\!\left(
        \mathcal A_j\frac{\Delta t_{\rm ctrl}}{2}
    \right)
    \mathbf y_j.
\end{equation}
Because the final trap is stationary, the endpoint displacement and velocity are
\begin{equation}
    \mathbf r_{\rm com}(T)
    =
    \bar{\mathbf r}(T)-\mathbf r_{\min}(T),
    \qquad
    \dot{\mathbf r}_{\rm com}(T)
    =
    \frac{\bar{\mathbf p}(T)}{m}
    -
    \dot{\mathbf r}_{\min}(T)
    =
    \frac{\bar{\mathbf p}(T)}{m}.
\end{equation}
The endpoint-fidelity estimate (Appendix~\ref{appendix:gaussian-propagation}) is then
\begin{equation}
    F_{\rm COM}
    =
    \exp\!\left[
        -\sum_{k=1}^{3}
        \left\{
            \left(\frac{r_{{\rm com},k}(T)}{2\sigma_{{\rm TF},k}}\right)^2
            +
            \left(\frac{m\,\sigma_{{\rm TF},k}\,\dot r_{{\rm com},k}(T)}{\hbar}\right)^2
        \right\}
    \right],
    \label{eq:com-overlap}
\end{equation}
where $k$ indexes the final trap principal axes and $\sigma_{{\rm TF},k}=R_{{\rm TF},k}(T)/\sqrt{7}$ is the equilibrium Thomas--Fermi rms width along axis $k$ (Appendix~\ref{appendix:tf-equilibrium}). Since $\sigma_{{\rm TF},k}$ increases with $N$, this estimate becomes more sensitive to residual velocity at larger atom numbers.

\subsection{Preflight Fidelity Scan}\label{sec:surrogate-screening}

\begin{figure}[!htbp]
    \centering
    \includegraphics[width=\textwidth]{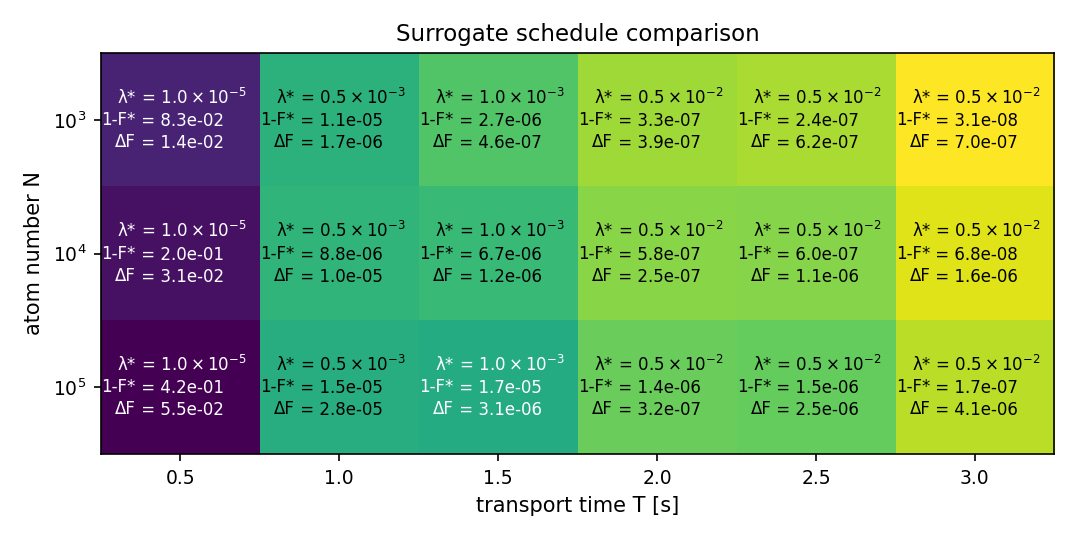}
    \caption{Surrogate schedule comparison for each $(N,T)$. Each cell reports the regularization $\lambda^*$ with the largest endpoint-fidelity estimate, its loss $1-F_{\rm COM}^*$, and its separation $\Delta F$ from the runner-up. Brighter cells indicate higher estimated overlap.}
    \label{fig:screening-selection}
\end{figure}

We evaluate $F_{\rm COM}$ for 18 $(N,T)$ pairs spanning $N\in\{10^3,10^4,10^5\}$ and six transport durations. The maximizing regularization within the sampled set $\Lambda$ is
\begin{equation}
    \lambda^*(N,T)
    =
    \underset{\lambda\in\Lambda}{\arg\max}\,
    F_{\rm COM}(\lambda;N,T),
    \qquad
    F_{\rm COM}^{*}(N,T)
    =
    F_{\rm COM}
    \left(
        \lambda^*(N,T);
        N,T
    \right).
    \label{eq:reg-selection}
\end{equation}
Fig.~\ref{fig:screening-selection} shows that the estimated endpoint fidelities are generally higher for lower atom numbers and longer transports. At $T=\SI{0.5}{\second}$, the winner exceeds its runner-up by $\Delta F=1.4\times10^{-2}$--$5.5\times10^{-2}$, whereas $\Delta F\leq2.8\times10^{-5}$ for $T\geq\SI{1.0}{\second}$. The same regularization maximizes the estimate at each atom number,
\begin{equation*}
    \lambda^*(T)
    =
    \begin{cases}
        1.0 \times 10^{-5}, & T=\SI{0.5}{\second},\\
        0.5 \times 10^{-3}, & T=\SI{1.0}{\second},\\
        1.0 \times 10^{-3}, & T=\SI{1.5}{\second},\\
        0.5 \times 10^{-2}, & T\in\{2.0,2.5,3.0\}\,\si{\second}.
    \end{cases}
\end{equation*}
At the fixed control cadence, shorter transports require larger node-to-node trap displacements. Large $\lambda$ suppresses the current updates needed to track them, so the fastest transport favors weaker regularization. Longer durations distribute the translation over more intervals and drive the trap more slowly, allowing stronger regularization with less residual COM excitation.

\begin{figure}[!htbp]
    \centering
    \includegraphics[width=\textwidth]{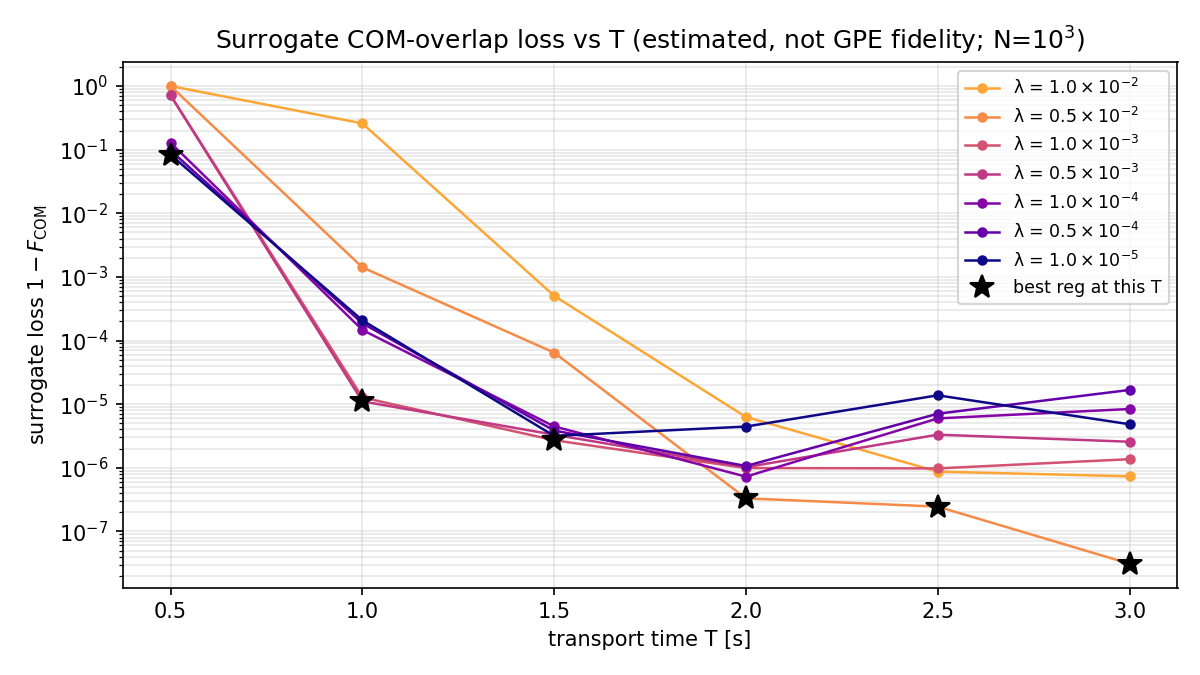}
    \caption{Estimated endpoint-fidelity loss $1-F_{\rm COM}$ for the seven candidate regularizations at $N=10^3$. Stars mark the smallest loss at each transport duration. These are estimates, not delivered GPE fidelities.}
    \label{fig:screening-overlap}
\end{figure}

Fig.~\ref{fig:screening-overlap} provides a complementary view of the same scan at $N=10^3$, showing the losses of all seven candidates on a logarithmic scale. This makes the small differences among candidates more visible and also reveals the shift from weaker regularization at shorter durations to stronger regularization at longer durations.

\subsection{COM Excursion During Transport}
\label{sec:com-excursion-in-transport}

Because endpoint fidelity does not constrain the motion during transport, we compare the maximum axial COM excursion with the cloud size by defining
\begin{equation}
    C_{\lambda,N}(T)
    =
    \frac{\max_j\left|r_{{\rm com},x}(t_j)\right|}
        {\sigma_{x,0}(N)},
    \qquad
    t_j\in[0,T],
    \label{eq:peak-slosh}
\end{equation}
where $\sigma_{x,0}(N)$ is the initial equilibrium Thomas--Fermi rms width projected onto the laboratory $x$ axis. Thus, $C_{\lambda,N}(T)=1$ indicates an excursion equal to one rms cloud width.

\begin{figure}[!htbp]
    \centering
    \includegraphics[width=\textwidth]{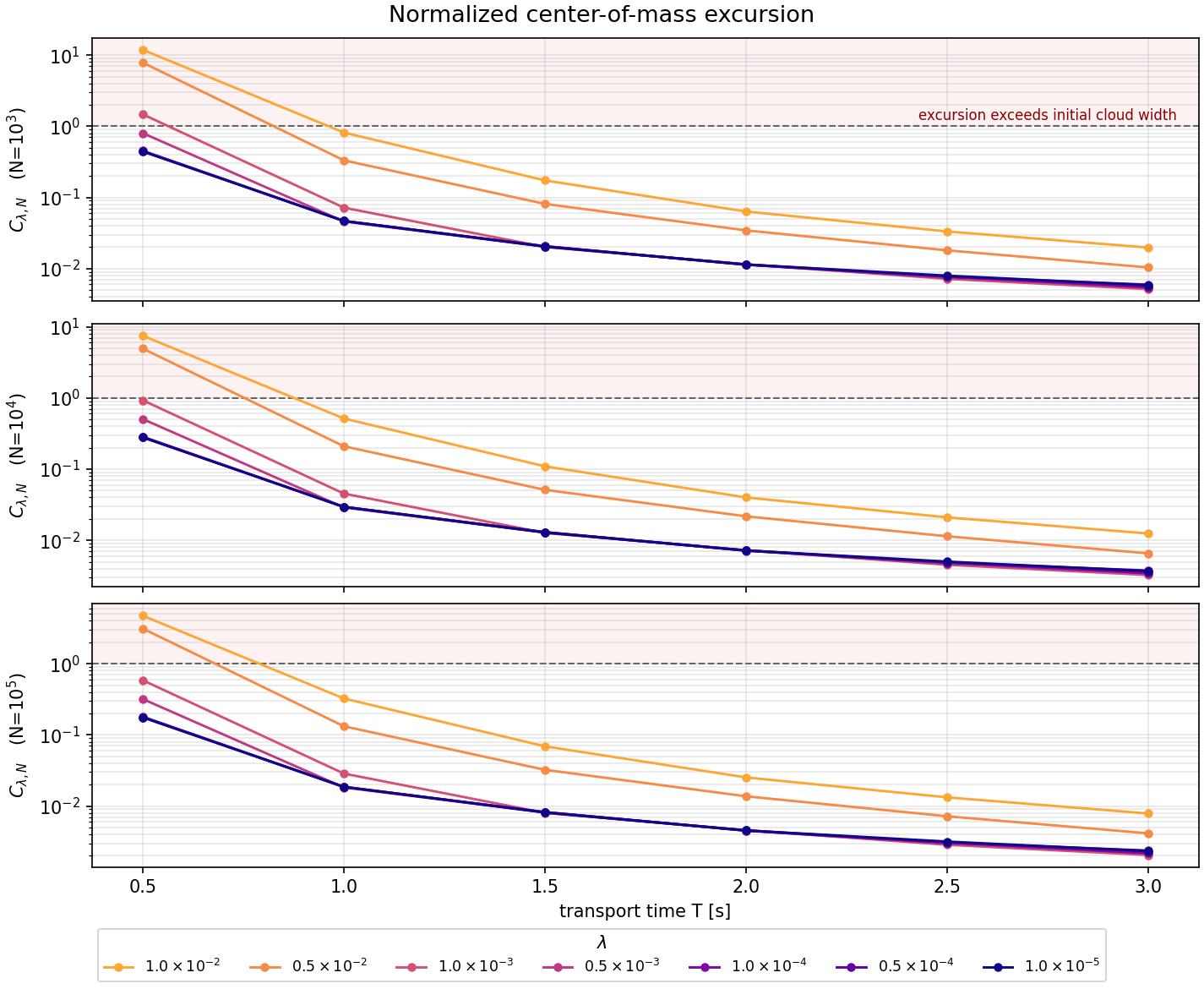}
    \caption{Maximum axial COM excursion over the control nodes, normalized by the initial equilibrium Thomas--Fermi rms width. Each row shows $C_{\lambda,N}(T)$ for the indicated atom number, with colors denoting the regularization. The dashed line marks an excursion equal to one cloud width.}
    \label{fig:transient-excursion}
\end{figure}

Fig.~\ref{fig:transient-excursion} shows that the maximum excursion decreases strongly with $T$, indicating closer COM tracking of the trap minimum at longer durations across the tested regularizations. The two largest regularizations produce substantially greater excursions, whereas the curves for $\lambda\leq10^{-4}$ nearly coincide, indicating that weaker regularization does not reduce COM motion further.
Because the Gaussian COM trajectory is independent of $N$, the smaller normalized values at larger atom numbers reflect the increasing Thomas--Fermi width rather than a change in the predicted motion. Every selected schedule satisfies $C_{\lambda,N}(T)<1$, so none reaches the reference value of one initial rms width. The COM-excursion analysis did not indicate a need to modify the endpoint-fidelity ranking for the present scan.

\section{Surrogate-Assisted Discretization}
\label{sec:surrogate-assisted-discretization}

We determine the case-specific numerical grid and time steps for each selected schedule.

\subsection{Grid Construction}

The spatial grid must be large enough to contain the predicted cloud motion, while its spacing must be small enough to resolve the finest estimated length scale.

\paragraph{Thomas--Fermi Scaling Model.}
We estimate changes in cloud size, orientation, and density using the Ermakov scaling approach~\cite{castin1996bose,kagan1996evolution}. Under harmonic confinement about the trap minimum and in the Thomas--Fermi hydrodynamic limit, the density retains a self-similar profile described by the scaling matrix $B(t)$,
\begin{equation}
    \ddot B(t)
    =
    -K(t)B(t)
    +
    \frac{1}{\det B(t)}B(t)^{-\top}K_0,
    \qquad
    B(0)=\mathbb I,
    \quad
    \dot B(0)=0,
    \label{eq:ermakov-matrix}
\end{equation}
where $K(t)=H(t)/m$ is the mass-normalized Hessian and $K_0=K(0)$ is its initial value (Appendix~\ref{appendix:tf-scaling}). We use the node-centered current convention (Eq.~\ref{eq:node-centered-current}) for $K(t)$.

\paragraph{Grid Dimensions.}
We construct the grid with equal extent on both sides of the trap minimum. Its half-width combines four contributions along each axis,
\begin{equation}
    \frac{L_i}{2}
    =
    E_i^{\rm Erm}
    +
    S_i^{\rm COM}
    +
    D_i^{\rm shift}
    +
    M_{\rm tail},
    \qquad
    i\in\{x,y,z\},
    \label{eq:box-sizing}
\end{equation}
where $E_i^{\rm Erm}$ is the maximum cloud half-extent along laboratory axis $i$ estimated by the Ermakov model, and $S_i^{\rm COM}$ is the maximum COM excursion relative to the trap minimum predicted by the Gaussian model. Because the grid is recentered at each midpoint current switch, the translation buffer is given by
\begin{equation}
    D_i^{\rm shift}
    =
    \max_j
    \left|
        r_{\min,i}(t_{j+1})
        -
        r_{\min,i}(t_j)
    \right|.
\end{equation}
The GPE density remains nonzero outside the finite Thomas--Fermi boundary, so we add a tail margin $M_{\rm tail}$ to the grid. We use the initial central healing length $\xi_0$ (Appendix~\ref{appendix:healing-length}) as the reference amplitude-variation scale. A general exponential sizing model represents the amplitude-suppression factor at distance $d$ by $\exp(-A_{\rm tail}d/\xi_0)$, where $A_{\rm tail}>0$ is a dimensionless decay-rate parameter. Because $n=|\psi|^2$, we define $M_{\rm tail}$ by setting the modeled relative-density factor at the grid margin equal to a dimensionless threshold $f_{\rm tail}$,
\begin{equation}
    \exp\!\left(
        -\frac{2A_{\rm tail}M_{\rm tail}}{\xi_0}
    \right)
    =
    \left[
        \exp\!\left(
            -\frac{2M_{\rm tail}}{\xi_0}
        \right)
    \right]^{A_{\rm tail}}
    =
    f_{\rm tail}.
\end{equation}
Defining $s_{\rm tail}=f_{\rm tail}^{1/A_{\rm tail}}$ absorbs the unknown decay-rate factor into a single effective parameter, giving
\begin{equation}
    \exp\!\left(
        -\frac{2M_{\rm tail}}{\xi_0}
    \right)
    =
    s_{\rm tail},
    \qquad
    M_{\rm tail}
    =
    \frac{\xi_0}{2}
    \ln\!\left(
        \frac{1}{s_{\rm tail}}
    \right).
    \label{eq:tail-margin}
\end{equation}
We choose the heuristic value $s_{\rm tail}=10^{-6}$ as a small effective suppression factor at the grid margin.

Table~\ref{tab:grid-dimensions} summarizes the $x$-axis half-width budget and the grid-box dimensions.
\begin{table}[!htbp]
    \centering
    \small
    \setlength{\tabcolsep}{6pt}
    \begin{tabular}{ccccccccc}
        \hline\\
        $N$
        & $T$ (\si{\second})
        & $\lambda^*$
        & $\xi_0$
        & $E_x^{\rm Erm}$
        & $S_x^{\rm COM}$
        & $D_x^{\rm shift}$
        & $M_{\rm tail}$
        & $L_x\times L_y\times L_z$ \\[2ex]
        \hline\\
        $10^3$ & 0.5 & $1\times10^{-5}$ & 0.27 & 4.50 & 0.59 & 4.52 & 1.88 & $22.98\times9.02\times7.34$ \\[1ex]
        $10^3$ & 1.0 & $5\times10^{-4}$ & 0.27 & 4.43 & 0.06 & 2.30 & 1.88 & $17.32\times8.51\times7.01$ \\[1ex]
        $10^3$ & 1.5 & $1\times10^{-3}$ & 0.27 & 4.39 & 0.03 & 1.54 & 1.88 & $15.67\times8.43\times6.92$ \\[1ex]
        $10^3$ & 2.0 & $5\times10^{-3}$ & 0.27 & 4.44 & 0.05 & 1.26 & 1.88 & $15.24\times8.43\times6.86$ \\[1ex]
        $10^3$ & 2.5 & $5\times10^{-3}$ & 0.27 & 4.41 & 0.02 & 0.98 & 1.88 & $14.58\times8.39\times6.85$ \\[1ex]
        $10^3$ & 3.0 & $5\times10^{-3}$ & 0.27 & 4.39 & 0.01 & 0.81 & 1.88 & $14.16\times8.37\times6.85$ \\[1ex]
        $10^4$ & 2.0 & $5\times10^{-3}$ & 0.17 & 7.04 & 0.05 & 1.26 & 1.18 & $19.05\times9.69\times7.23$ \\[2ex]
        \hline
    \end{tabular}
    \caption{Grid-dimension budgets and grid-box dimensions for the selected GPE evaluations. The healing length and four budget terms are reported in \si{\micro\meter}; the budget terms sum to $L_x/2$, and the final column gives the full grid lengths in \si{\micro\meter}. Here, $\lambda^*$ denotes the surrogate-selected regularization.}
    \label{tab:grid-dimensions}
\end{table}
For $N=10^3$, the translation buffer $D_x^{\rm shift}$ is the main source of variation in $L_x$ and generally decreases across the selected schedules because, at the fixed control interval $\Delta t_{\rm ctrl}$, longer-duration transports contain more control nodes over which the trap motion is distributed. The general decrease in $S_x^{\rm COM}$ follows the surrogate COM trend discussed in Section~\ref{sec:com-excursion-in-transport}. By contrast, $E_x^{\rm Erm}$ remains near $4.4\,\mu\mathrm{m}$, and $M_{\rm tail}$ remains fixed because the $N=10^3$ cases share the same $\xi_0$ and $s_{\rm tail}$.

For $N=10^4$, comparison with the $N=10^3$ case (the same $T=2.0\,\mathrm{s}$ and $\lambda^*=5\times10^{-3}$) shows that $E_x^{\rm Erm}$ increases from $4.44$ to $7.04\,\mu\mathrm{m}$ because of the larger interaction-dependent cloud extent. This increase is consistent with the Thomas–Fermi scaling $E_x^{\rm Erm}\propto N^{1/5}$ (Appendix~\ref{appendix:tf-scaling}). The higher chemical potential at $N=10^4$ reduces $\xi_0$ and hence $M_{\rm tail}$.

\paragraph{Grid Resolution.}
The grid resolution is based on the healing length $\xi$, which characterizes the spatial scale at which kinetic effects from amplitude variations become non-negligible. The Ermakov model estimates its minimum value $\xi_{\min}$ during transport (Appendix~\ref{appendix:healing-length}). We use a resolution criterion of at least two grid intervals across $\xi_{\min}$ and select $N_i$ such that
\begin{equation}
    \Delta r_i
    =
    \frac{L_i}{N_i}
    \leq
    \frac{\xi_{\min}}{2},
    \qquad
    i\in\{x,y,z\}.
    \label{eq:spatial-resolution}
\end{equation}
Once the grid lengths $L_i$ and target spacing $\xi_{\min}/2$ have been obtained, we choose $N_i$ as the smallest integer containing only the prime factors $2$, $3$, $5$, and $7$ that satisfies $N_i\geq2L_i/\xi_{\min}$, thereby using the transform sizes for which cuFFT provides optimized algorithms~\cite{nvidia2025cufft}.

Table~\ref{tab:grid-sizing} summarizes the cell counts and spatial resolutions. 
\begin{table}[!htbp]
    \centering
    \small
    \setlength{\tabcolsep}{6pt}
    \begin{tabular}{cccccc}
        \hline\\
        $N$
        & $T$
        & $\lambda^*$
        & $\xi_{\min}$
        & $N_x\times N_y\times N_z$
        & $\Delta r_x\times\Delta r_y\times\Delta r_z$ \\
        & (\si{\second})
        &
        & (\si{\nano\meter})
        &
        & (\si{\nano\meter}) \\[2ex]
        \hline\\
        $10^3$ & 0.5 & $1\times10^{-5}$ & 251.5 & $189\times72\times60$ & $121.6\times125.2\times122.3$ \\[1ex]
        $10^3$ & 1.0 & $5\times10^{-4}$ & 252.3 & $140\times70\times56$ & $123.7\times121.5\times125.1$ \\[1ex]
        $10^3$ & 1.5 & $1\times10^{-3}$ & 252.5 & $125\times70\times56$ & $125.4\times120.4\times123.6$ \\[1ex]
        $10^3$ & 2.0 & $5\times10^{-3}$ & 253.7 & $125\times70\times56$ & $121.9\times120.4\times122.6$ \\[1ex]
        $10^3$ & 2.5 & $5\times10^{-3}$ & 253.3 & $120\times70\times56$ & $121.5\times119.8\times122.3$ \\[1ex]
        $10^3$ & 3.0 & $5\times10^{-3}$ & 253.1 & $112\times70\times56$ & $126.5\times119.6\times122.2$ \\[1ex]
        $10^4$ & 2.0 & $5\times10^{-3}$ & 160.1 & $240\times125\times96$ & $79.4\times77.5\times75.3$ \\[2ex]
        \hline
    \end{tabular}
    \caption{Realized cell counts and spatial resolutions for the selected GPE evaluations, where $\lambda^*$ denotes the surrogate-selected regularization. Each triplet follows the $(x,y,z)$ axis order.}
    \label{tab:grid-sizing}
\end{table}
For $N=10^3$, $\xi_{\min}$ varies by less than $1\%$ across the six selected cases, so the differences in cell counts mainly reflect changes in the grid lengths. The spacings are approximately $94$--$100\%$ of $\xi_{\min}/2$, showing that the FFT-friendly cell counts satisfy the resolution criterion with little additional refinement. At $N=10^4$, the smaller healing length requires spacings near $0.08\,\mu\mathrm{m}$ and substantially larger cell counts.

\subsection{Time-Step Selection}

We next determine the parameters for real-time propagation and imaginary-time relaxation.

\paragraph{Real-Time Propagation Step.}
In the spatial grid constructed above, the shortest distinguishable wavelength along axis $i$ spans two grid intervals, $2\Delta r_i$.
The upper limit on the representable magnitude of the wavevector component is therefore
\begin{equation}
    k_{{\rm Nyq},i}
    =
    \frac{2\pi}{2\Delta r_i}
    =
    \frac{\pi}{\Delta r_i},
\end{equation}
which is called the Nyquist limit along axis $i$. Taking all three wavevector components at their Nyquist limits gives the maximum kinetic phase rate represented on the grid,
\begin{equation}
    \omega_{\rm kin}^{\rm Nyq}
    =
    \frac{\hbar}{2m}
    \sum_{i\in\{x,y,z\}}
    k_{{\rm Nyq},i}^{2}
    =
    \frac{\hbar}{2m}
    \sum_{i\in\{x,y,z\}}
    \left(
        \frac{\pi}{\Delta r_i}
    \right)^2.
\end{equation}
We also include the maximum trap frequency $\omega_{\rm trap}^{\max}$ as a heuristic allowance for trap dynamics to select the time step.
\begin{equation}
    \left(
        \omega_{\rm kin}^{\rm Nyq}
        +
        \omega_{\rm trap}^{\max}
    \right)
    \Delta t
    \leq
    \phi_{\rm target},
    \qquad
    \omega_{\rm trap}^{\max}
    =
    \max_{0\leq t\leq T}
    \max_{i\in\{1,2,3\}}
    \omega_i(t).
\end{equation}
We use a phase increment of $\phi_{\rm target}=0.4$ per step, which is approximately $16$ steps per phase cycle. We round the upper bound on $\Delta t$ downward to a multiple of $0.1\,\mu\mathrm{s}$, giving the provisional maximum $\Delta t_{\max}$. We then choose $n_{\rm sub}\in2\mathbb N$ such that the realized propagation step satisfies
\begin{equation}
    \Delta t
    =
    \frac{\Delta t_{\rm ctrl}}{n_{\rm sub}}
    \leq
    \Delta t_{\max}.
\end{equation}

Table~\ref{tab:phase-rates} lists the phase rates and calculated time steps.
\begin{table}[!htbp]
    \centering
    \small
    \setlength{\tabcolsep}{6pt}
    \begin{tabular}{ccccccccc}
        \hline\\
        $N$
        & $T$ (\si{\second})
        & $\lambda^*$
        & $\omega_{\rm kin}^{\rm Nyq}$ (\si{\radian\per\second})
        & $\omega_{\rm trap}^{\max}$ (\si{\radian\per\second})
        & $\dfrac{\omega_{\rm trap}^{\max}}{\omega_{\rm kin}^{\rm Nyq}}$ (\si{\percent})
        & $\Delta t_{\max}$ (\si{\micro\second})
        & $\Delta t$ (\si{\micro\second})
        & $n_{\rm sub}$ \\[2ex]
        \hline\\
        $10^3$ & 0.5 & $1\times10^{-5}$ & $7.149\times10^5$ & $2.717\times10^3$ & 0.3800 & 0.5 & 0.5 & 1000 \\[1ex]
        $10^3$ & 1.0 & $5\times10^{-4}$ & $7.104\times10^5$ & $2.695\times10^3$ & 0.3793 & 0.5 & 0.5 & 1000 \\[1ex]
        $10^3$ & 1.5 & $1\times10^{-3}$ & $7.143\times10^5$ & $2.690\times10^3$ & 0.3765 & 0.5 & 0.5 & 1000 \\[1ex]
        $10^3$ & 2.0 & $5\times10^{-3}$ & $7.316\times10^5$ & $2.657\times10^3$ & 0.3633 & 0.5 & 0.5 & 1000 \\[1ex]
        $10^3$ & 2.5 & $5\times10^{-3}$ & $7.363\times10^5$ & $2.667\times10^3$ & 0.3623 & 0.5 & 0.5 & 1000 \\[1ex]
        $10^3$ & 3.0 & $5\times10^{-3}$ & $7.191\times10^5$ & $2.673\times10^3$ & 0.3717 & 0.5 & 0.5 & 1000 \\[1ex]
        $10^4$ & 2.0 & $5\times10^{-3}$ & $1.808\times10^6$ & $2.657\times10^3$ & 0.1470 & 0.2 & 0.2 & 2500 \\[2ex]
        \hline
    \end{tabular}
    \caption{Nyquist kinetic and maximum trap phase rates and the resulting real-time propagation parameters for the selected GPE evaluations. In every case, the realized step $\Delta t$ equals the rounded provisional maximum $\Delta t_{\max}$.}
    \label{tab:phase-rates}
\end{table}
 For $N=10^3$, $\omega_{\rm trap}^{\max}$ is only $0.36$--$0.38\%$ of $\omega_{\rm kin}^{\rm Nyq}$ across all six selected cases, so the kinetic Nyquist rate dominates the time-step bound. Adding the trap contribution therefore does not change the rounded $\Delta t_{\max}$, which remains $0.5\,\mu\mathrm{s}$ in every case. Each control interval used $n_{\rm sub}=1000$ propagation steps, giving $\Delta t=\SI{0.5}{\micro\second}$.

The $N=10^4$, $T=\SI{2.0}{\second}$ case requires a finer grid because of its smaller healing length. The resulting increase in $\omega_{\rm kin}^{\rm Nyq}$ reduces $\Delta t_{\max}$ to $\SI{0.2}{\micro\second}$. Choosing $n_{\rm sub}=2500$ per control interval then gives the realized step $\Delta t=\SI{0.2}{\micro\second}$. More propagation steps, along with the larger cell count, make the $N=10^4$ case substantially more computationally demanding.

\paragraph{Imaginary-Time Propagation Step.}
We use the trap frequencies as heuristic relaxation rates because harmonic-trap excitation gaps are of order $\hbar\omega_i$, giving imaginary-time decay rates of order $\omega_i$. We choose $\Delta\tau$ so that the decay exponent per iteration at the fastest trap scale is approximately $0.1$,
\begin{equation}
    \omega_{\rm trap}^{\max}\Delta\tau
    \simeq
    0.1.
\end{equation}
We round $\Delta\tau$ to the nearest \SI{0.5}{\micro\second}. We choose $n_{\rm imag}$ by requiring the estimated squared-amplitude decay factor at the slowest trap scale to be at most $10^{-6}$,
\begin{equation}
    \exp\!\left(
        -2\omega_{\rm trap}^{\min}
        n_{\rm imag}\Delta\tau
    \right)
    \leq
    10^{-6},
    \qquad
    \omega_{\rm trap}^{\min}
    =
    \min_{0\leq t\leq T}
    \min_{i\in\{1,2,3\}}
    \omega_i(t),
\end{equation}
where $\omega_{\rm trap}^{\min}$ is the slowest trap scale. We round $n_{\rm imag}$ upward to the nearest 100. Table~\ref{tab:imaginary-time-parameters} lists the trap-frequency scales and calculated relaxation parameters.
\begin{table}[!htbp]
    \centering
    \small
    \setlength{\tabcolsep}{6pt}
    \begin{tabular}{ccccccc}
        \hline\\
        $N$
        & $T$ (\si{\second})
        & $\lambda^*$
        & $\omega_{\rm trap}^{\min}$ (\si{\radian\per\second})
        & $\omega_{\rm trap}^{\max}$ (\si{\radian\per\second})
        & $\Delta\tau$ (\si{\micro\second})
        & $n_{\rm imag}$ \\[2ex]
        \hline\\
        $10^3$ & 0.5 & $1\times10^{-5}$ & $4.619\times10^2$ & $2.717\times10^3$ & 37.0 & 500 \\[1ex]
        $10^3$ & 1.0 & $5\times10^{-4}$ & $4.676\times10^2$ & $2.695\times10^3$ & 37.0 & 400 \\[1ex]
        $10^3$ & 1.5 & $1\times10^{-3}$ & $4.698\times10^2$ & $2.690\times10^3$ & 37.0 & 400 \\[1ex]
        $10^3$ & 2.0 & $5\times10^{-3}$ & $4.662\times10^2$ & $2.657\times10^3$ & 37.5 & 400 \\[1ex]
        $10^3$ & 2.5 & $5\times10^{-3}$ & $4.684\times10^2$ & $2.667\times10^3$ & 37.5 & 400 \\[1ex]
        $10^3$ & 3.0 & $5\times10^{-3}$ & $4.699\times10^2$ & $2.673\times10^3$ & 37.5 & 400 \\[1ex]
        $10^4$ & 2.0 & $5\times10^{-3}$ & $4.662\times10^2$ & $2.657\times10^3$ & 37.5 & 400 \\[2ex]
        \hline
    \end{tabular}
    \caption{Trap-frequency scales and imaginary-time relaxation parameters for the selected GPE evaluations.}
    \label{tab:imaginary-time-parameters}
\end{table}
The trap-frequency extrema vary by less than $3\%$. The slightly smaller $\omega_{\rm trap}^{\min}$ for $T=0.5\,\mathrm{s}$ raises $n_{\rm imag}$ to 500. This heuristic depends on the trap frequencies rather than $N$, so the two $T=2.0\,\mathrm{s}$ cases use identical relaxation parameters.

\paragraph{Relaxation Residual.}
After imaginary-time relaxation to find the ground state at $t\in\{0,T\}$, each relaxed wavefunction should satisfy the stationary GPE,
\begin{equation}
    (\hat T+\hat V)\psi
    =
    \mu\psi,
    \qquad
    \mu
    =
    \int
    \psi^*(\hat T+\hat V)\psi
    \,d^3\mathbf r,
\end{equation}
where $\mu$ is the chemical-potential. With this definition, any remaining component can be written as
\begin{equation}
    (\hat T+\hat V)\psi
    =
    \mu\psi+\chi_\perp,
\end{equation}
where $\chi_\perp$ is orthogonal to $\psi$. We quantify its magnitude using the dimensionless residual
\begin{equation}
    R_{\rm stat}
    =
    \frac{
        \sqrt{
            \int
            \left|(\hat T+\hat V-\mu)\psi\right|^2
            \,d^3\mathbf r
        }
    }{
        |\mu|
    }.
    \label{eq:stationary-residual}
\end{equation}
For every selected GPE calculation, the initial and target stationary-state residuals remain below $10^{-5}$.

\section{GPE Characterization}

We characterize the endpoint fidelity and condensate dynamics for each surrogate-selected schedule by GPE propagation. For comparison, we also evaluate each case with $g_{\rm eff}=0$ as a linear baseline.

\subsection{Endpoint Fidelity Performance}

Table~\ref{tab:gpe-fidelity} compares their endpoint losses with the surrogate estimate at $N=10^3$. We report $1-F$ so that differences among near-unit fidelities remain visible.
\begin{table}[!htbp]
    \centering
    \small
    \setlength{\tabcolsep}{6pt}
    \begin{tabular}{ccccc}
        \hline\\
        $T$ (\si{\second})
        & $\lambda$
        & $1-F_{\rm COM}$
        & $1-F_{3D}$
        & $1-F_{\rm linear}$ \\[2ex]
        \hline\\
        0.5 & $1.0\times10^{-5}$          & $8.31\times10^{-2}$ & $7.56\times10^{-2}$ & $2.11\times10^{-2}$ \\[1ex]
        1.0 & $0.5\times10^{-3}$   & $1.10\times10^{-5}$ & $2.38\times10^{-4}$ & $3.97\times10^{-5}$ \\[1ex]
        1.5 & $1.0\times10^{-3}$          & $2.71\times10^{-6}$ & $2.23\times10^{-4}$ & $1.11\times10^{-6}$ \\[1ex]
        2.0 & $0.5\times10^{-2}$   & $3.31\times10^{-7}$ & $1.23\times10^{-4}$ & $5.77\times10^{-7}$ \\[1ex]
        2.5 & $0.5\times10^{-2}$   & $2.45\times10^{-7}$ & $1.93\times10^{-4}$ & $6.25\times10^{-7}$ \\[1ex]
        3.0 & $0.5\times10^{-2}$   & $3.10\times10^{-8}$ & $2.74\times10^{-4}$ & $4.31\times10^{-7}$ \\[2ex]
        \hline
    \end{tabular}
    \caption{Endpoint losses for the surrogate-selected schedules at $N=10^3$.}
    \label{tab:gpe-fidelity}
\end{table}
Across all tested durations, the surrogate is qualitatively consistent with the GPE. At $T=\SI{0.5}{\second}$, both indicate appreciable endpoint excitation, with losses $1-F_{\rm COM}=8.31\times10^{-2}$ and $1-F_{3D}=7.56\times10^{-2}$ for the selected schedule. For the longer transports, both indicate near-unit delivery. Every selected schedule with $T\geq\SI{1.0}{\second}$ gives $1-F_{3D}<3\times10^{-4}$ under GPE propagation. For these longer transports, $1-F_{\rm COM}$ and $1-F_{\rm linear}$ are generally similar in magnitude and both much smaller than $1-F_{3D}$, indicating that the residual GPE loss primarily reflects interaction-dependent internal dynamics absent from the COM surrogate and linear calculation.

\subsection{Condensate Dynamics During Transport}

\paragraph{COM Excursion.}
Table~\ref{tab:com-validation} shows that the surrogate and GPE excursions agree closely across all six schedules at $N=10^3$, with a maximum absolute difference of \SI{5.6}{\nano\meter}.
\begin{table}[!htbp]
    \centering
    \small
    \setlength{\tabcolsep}{6pt}
    \begin{tabular}{cccc}
        \multicolumn{4}{c}{Maximum $|r_{{\rm com},x}|$ (\si{\micro\meter})} \\[1ex]
        \hline\\
        $T$ (\si{\second})
        & Surrogate
        & GPE
        & Difference \\[2ex]
        \hline\\
        0.5 & 0.5932 & 0.5988 & 0.0056 \\[1ex]
        1.0 & 0.0614 & 0.0620 & 0.0006 \\[1ex]
        1.5 & 0.0269 & 0.0272 & 0.0003 \\[1ex]
        2.0 & 0.0461 & 0.0450 & 0.0011 \\[1ex]
        2.5 & 0.0242 & 0.0234 & 0.0008 \\[1ex]
        3.0 & 0.0139 & 0.0131 & 0.0008 \\[2ex]
        \hline
    \end{tabular}
    \caption{Maximum sampled axial COM excursion for the surrogate-selected schedules at $N=10^3$: Gaussian phase-space predictions and GPE simulation results, both evaluated at the control-node times.}
    \label{tab:com-validation}
\end{table}

\begin{figure}[!htbp]
    \centering
    \includegraphics[width=0.9\textwidth]{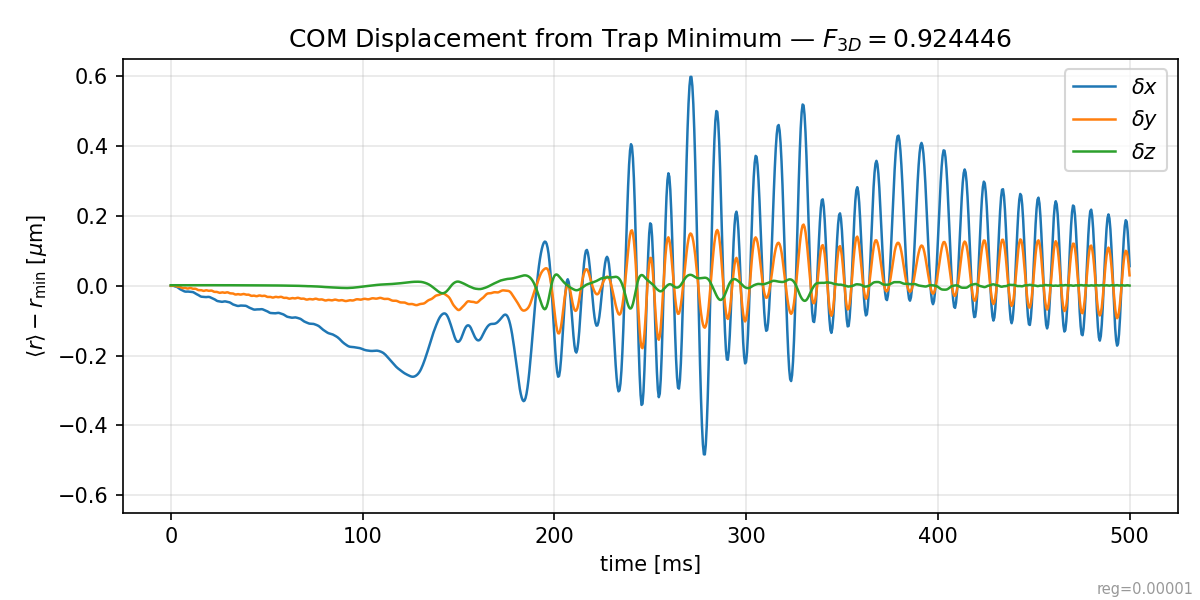}
    \caption{COM displacement from the trap minimum for $N=10^3$, $T=\SI{0.5}{\second}$, and $\lambda=10^{-5}$.}
    \label{fig:gpe-com-fast}
\end{figure}

\paragraph{COM Sloshing.}
The shortest selected transport, which also uses the weakest regularization, produces the clearest COM excitation. At $T=\SI{0.5}{\second}$ and $\lambda=10^{-5}$, the axial COM displacement reaches approximately \SI{0.60}{\micro\meter} (maximum $|\mathbf r_{\rm com}|$ of \SI{0.62}{\micro\meter}), and the oscillations persist through delivery (Fig.~\ref{fig:gpe-com-fast}). The endpoint fidelity is $F_{3D}=0.9244$.

\begin{figure}[!htbp]
    \centering
    \includegraphics[width=0.9\textwidth]{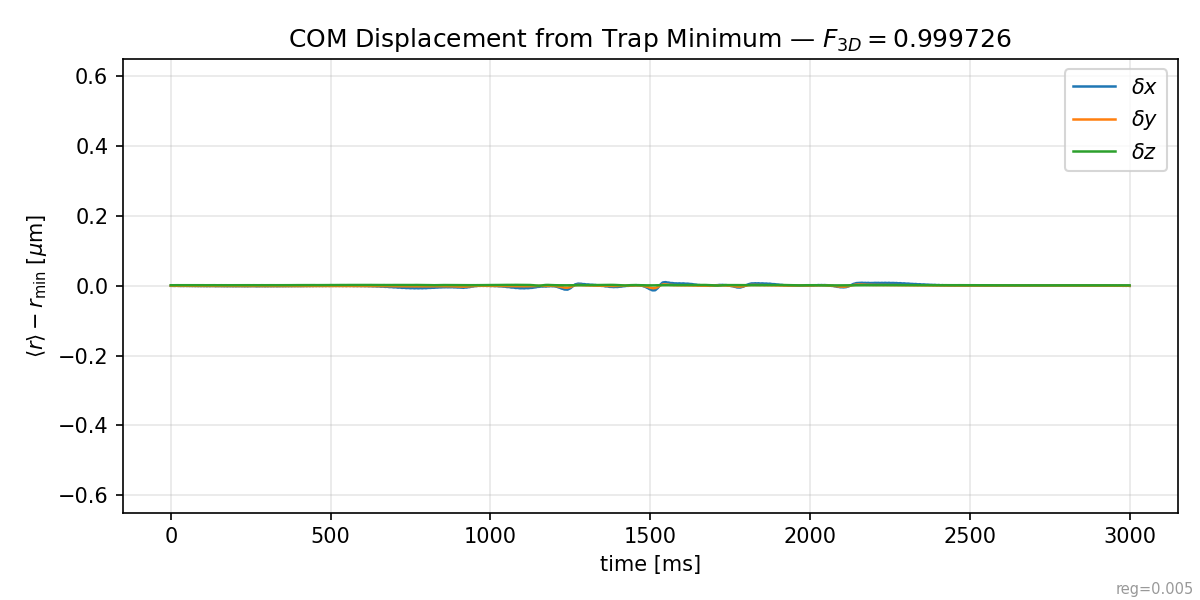}
    \caption{COM displacement from the trap minimum for $N=10^3$, $T=\SI{3.0}{\second}$, and $\lambda=5\times10^{-3}$.}
    \label{fig:gpe-com-slow}
\end{figure}

In contrast, at $T=\SI{3.0}{\second}$ and $\lambda=5\times10^{-3}$, the axial COM displacement reaches approximately \SI{0.013}{\micro\meter} (maximum $|\mathbf r_{\rm com}|$ of \SI{0.015}{\micro\meter}) (Fig.~\ref{fig:gpe-com-slow}). The endpoint fidelity is $F_{3D}=0.999726$.

For the two schedules shown, the trap minimum moves transversely by several micrometres, while the COM displacement relative to it remains submicrometre.

\begin{figure}[!htbp]
    \centering
    \includegraphics[width=0.9\textwidth]{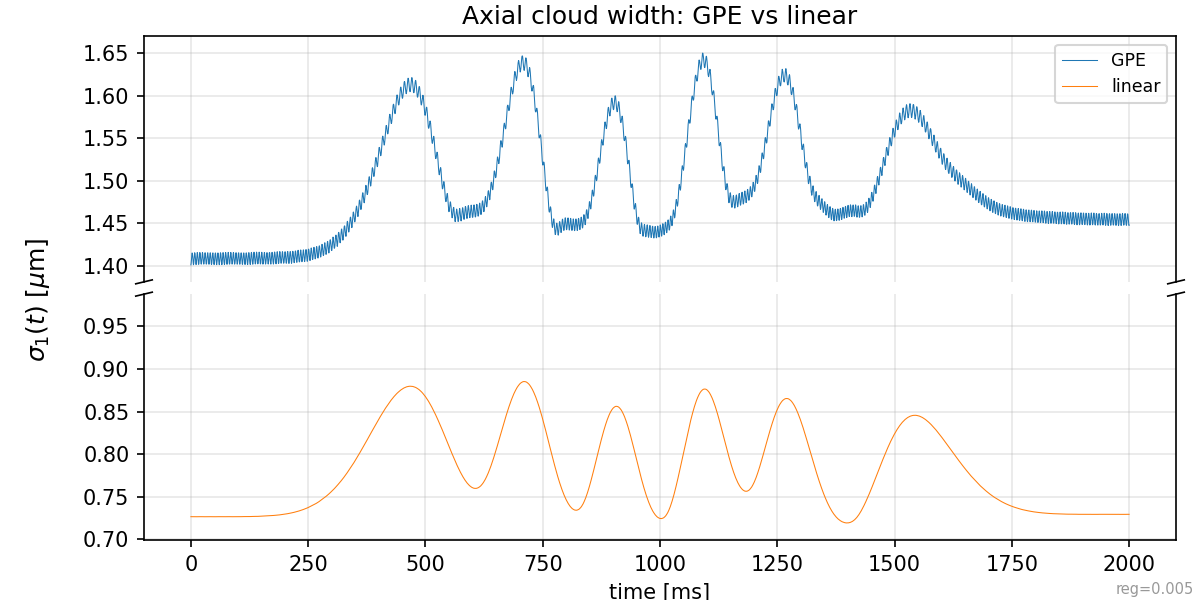}
    \caption{Axial cloud width under GPE and linear evolution for $N=10^3$, $T=\SI{2.0}{\second}$, and $\lambda=5\times10^{-3}$.}
    \label{fig:gpe-width-slow}
\end{figure}

\paragraph{Axial Width Dynamics.}
Fig.~\ref{fig:gpe-width-slow} compares the axial cloud widths from the GPE and linear $g_{\rm eff}=0$ calculations for $T=\SI{2.0}{\second}$.
Repulsive interatomic interactions broaden the initial GPE width to \SI{1.401}{\micro\meter}, roughly twice the linear value of \SI{0.727}{\micro\meter}. Under the same time-dependent external potential, both widths follow a similar overall temporal pattern, but the GPE has a larger absolute variation and exhibits superimposed fine oscillations.

\paragraph{Cloud Breathing.}
To account for changes in the trap confinement, we compare the GPE and Thomas--Fermi widths using
\begin{equation}
    b_k(t)
    =
    \frac{
        \sigma_k(t)/\sigma_k(0)
    }{
        \sigma_{{\rm TF},k}(t)/\sigma_{{\rm TF},k}(0)
    },
    \qquad
    k\in\{1,\perp\},
\end{equation}
where $k=1$ denotes the axial direction, while for $k=\perp$ we use the transverse geometric means
\begin{equation}
    \sigma_\perp
    =
    \sqrt{\sigma_2\sigma_3},
    \qquad
    \sigma_{{\rm TF},\perp}
    =
    \sqrt{\sigma_{{\rm TF},2}\sigma_{{\rm TF},3}}.
\end{equation}
$b_k(t)=1$ means that the cloud width changes in the same proportion as the Thomas--Fermi prediction. 

\begin{figure}[H]
    \centering
    \includegraphics[width=0.9\textwidth]{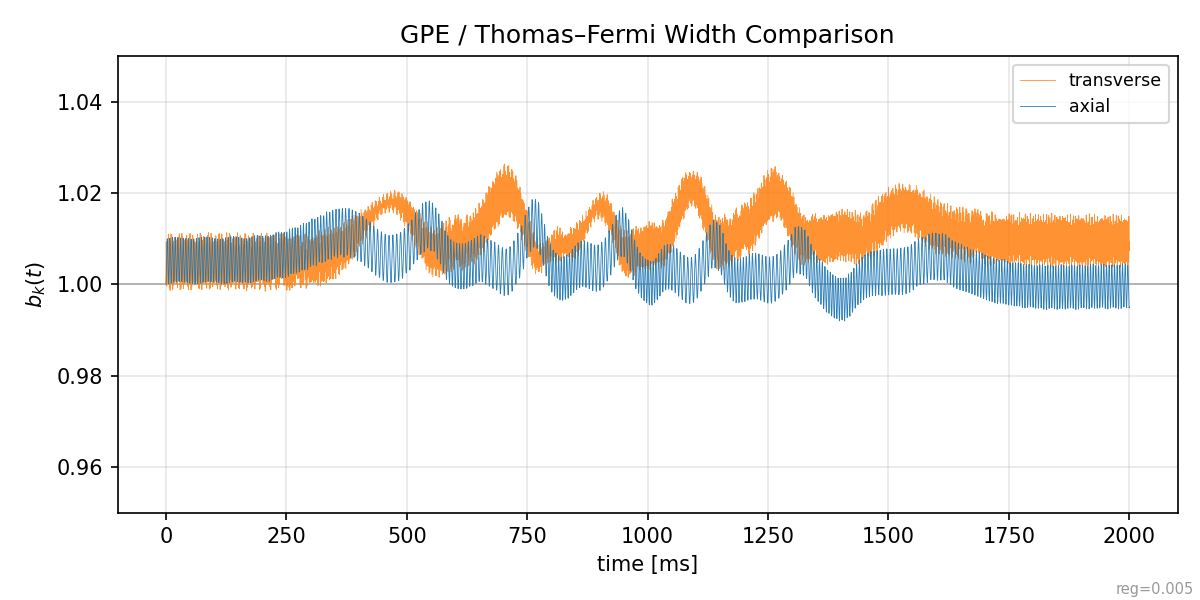}
    \caption{Ratio of the initial-normalized GPE and Thomas--Fermi widths for $N=10^3$, $T=\SI{2.0}{\second}$, and $\lambda=5\times10^{-3}$.}
    \label{fig:gpe-breathing-slow}
\end{figure}

Fig.~\ref{fig:gpe-breathing-slow} shows that the GPE widths broadly follow the Thomas--Fermi prediction, while also revealing superimposed axial and transverse oscillations.

\paragraph{Cloud Orientation.}
The orientation channel is quiet throughout. Across the selected schedules, the maximum axial cloud--trap misalignment $\theta_1(t)$ (Eq.~\ref{eq:gpe-misalignment}) decreases from $0.039^\circ$ at $T=\SI{0.5}{\second}$ to $0.0023^\circ$ at $T=\SI{3.0}{\second}$, so the transported cloud follows the axial confinement direction closely.

\subsection{10,000-Atom Case}
\label{sec:larger-n}

The $N=10^4$ GPE calculation uses a larger grid and more propagation steps per control interval (Tables~\ref{tab:grid-sizing} and \ref{tab:phase-rates}), which requires substantially more computation. We therefore evaluate it only at $T=\SI{2.0}{\second}$, which gives the highest GPE endpoint fidelity among the selected $N=10^3$ schedules. At this duration, the surrogate scan selects $\lambda=5\times10^{-3}$ at both atom numbers.

\paragraph{Endpoint Fidelity Comparison.}
Table~\ref{tab:atom-number-fidelity} compares the endpoint losses. The $N=10^4$ case has a lower GPE endpoint loss, while the linear loss is unchanged because setting $g_{\rm eff}=0$ removes the atom-number dependence from the evolution.
\begin{table}[!htbp]
    \centering
    \small
    \setlength{\tabcolsep}{6pt}
    \begin{tabular}{cccc}
        \hline\\
        $N$
        & $1-F_{\rm COM}$
        & $1-F_{3D}$
        & $1-F_{\rm linear}$ \\[2ex]
        \hline\\
        $10^3$
        & $3.31\times10^{-7}$
        & $1.23\times10^{-4}$
        & $5.77\times10^{-7}$ \\[1ex]
        $10^4$
        & $5.80\times10^{-7}$
        & $6.77\times10^{-5}$
        & $5.77\times10^{-7}$ \\[2ex]
        \hline
    \end{tabular}
    \caption{Endpoint losses at $T=\SI{2.0}{\second}$ and $\lambda=5\times10^{-3}$.}
    \label{tab:atom-number-fidelity}
\end{table}

\paragraph{COM Excursion Comparison.}
Table~\ref{tab:atom-number-com} compares the maximum axial COM excursions. The nearly unchanged GPE excursion shows that the lower endpoint loss at $N=10^4$ is not accompanied by a comparable change in COM motion.
\begin{table}[!htbp]
    \centering
    \small
    \setlength{\tabcolsep}{6pt}
    \begin{tabular}{ccc}
        \hline\\
        $N$
        & Gaussian (\si{\micro\meter})
        & GPE (\si{\micro\meter}) \\[2ex]
        \hline\\
        $10^3$ & 0.0461 & 0.0450 \\[1ex]
        $10^4$ & 0.0461 & 0.0441 \\[2ex]
        \hline
    \end{tabular}
    \caption{Maximum axial COM excursions at $T=\SI{2.0}{\second}$ and $\lambda=5\times10^{-3}$.}
    \label{tab:atom-number-com}
\end{table}

\paragraph{Cloud-Width Comparison.}
Table~\ref{tab:atom-number-widths} shows that the absolute axial and transverse widths are larger at $N=10^4$. 
\begin{table}[!htbp]
    \centering
    \small
    \setlength{\tabcolsep}{6pt}
    \begin{tabular}{ccccc@{\qquad}cccc}
        \hline\\
        & \multicolumn{4}{c@{\qquad}}{Axial $\sigma_1$ (\si{\micro\meter})}
        & \multicolumn{4}{c}{Transverse $\sigma_\perp$ (\si{\micro\meter})} \\[1ex]
        $N$
        & $t=0$ & Minimum & Maximum & $t=T$
        & $t=0$ & Minimum & Maximum & $t=T$ \\[2ex]
        \hline\\
        $10^3$
        & 1.401 & 1.401 & 1.650 & 1.448
        & 0.526 & 0.483 & 0.644 & 0.506 \\[1ex]
        $10^4$
        & 2.297 & 2.297 & 2.693 & 2.390
        & 0.752 & 0.689 & 0.912 & 0.718 \\[2ex]
        \hline
    \end{tabular}
    \caption{GPE axial and transverse rms widths at $T=\SI{2.0}{\second}$ and $\lambda=5\times10^{-3}$. Here, $\sigma_\perp=\sqrt{\sigma_2\sigma_3}$, and the extrema are evaluated at the recorded control-node times.}
    \label{tab:atom-number-widths}
\end{table}
For both cases, the cloud's axial and transverse widths vary over ranges of approximately $17\%$ and $30\%$ of their respective initial values, and the final cloud is more elongated than the initial cloud because the endpoint trap has slightly weaker axial confinement and stronger transverse confinement. The broad width variation and endpoint elongation reflect changes in confinement permitted by the inverse optimization, which constrains the trap-minimum trajectory but not the trap Hessian. The unconstrained frequencies could change in either direction, so preserving the confinement would require extending the inverse-optimization objective rather than modifying the GPE propagation.

\section{Numerical Adequacy}

We assess the numerical adequacy of the surrogate-assisted discretization described in Section~\ref{sec:surrogate-assisted-discretization}.

\subsection{Grid-Sizing Budget}

\begin{figure}[!htbp]
    \centering
    \begin{subfigure}{0.9\textwidth}
        \centering
        \includegraphics[width=\textwidth]{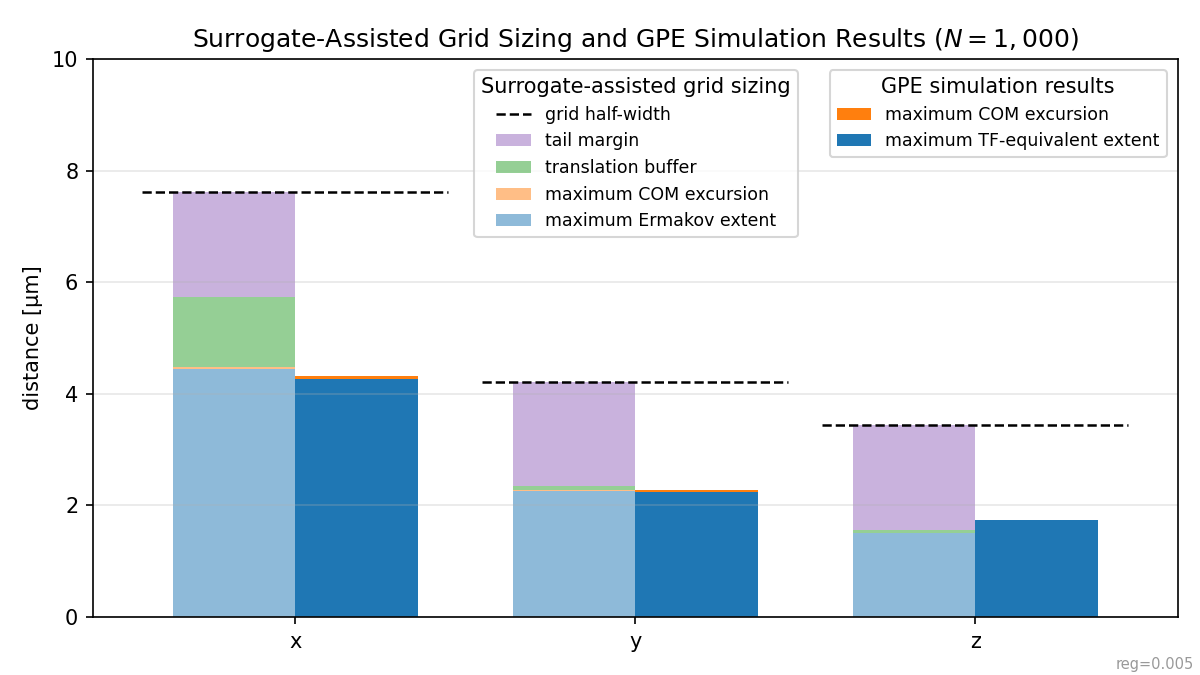}
    \end{subfigure}

    \smallskip

    \begin{subfigure}{0.9\textwidth}
        \centering
        \includegraphics[width=\textwidth]{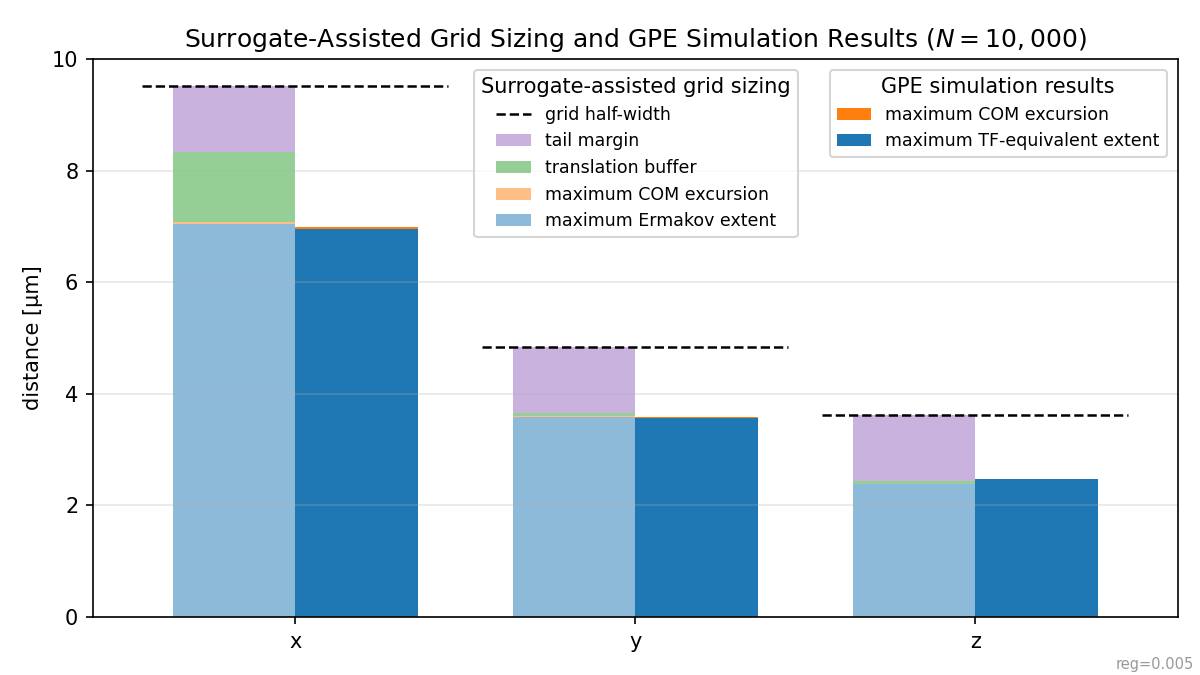}
    \end{subfigure}

    \caption{Surrogate-assisted grid sizing and GPE simulation results for the selected $T=\SI{2.0}{\second}$ cases.}
    \label{fig:sizer-adequacy}
\end{figure}

Figure~\ref{fig:sizer-adequacy} compares the surrogate-assisted grid sizing with the GPE results for $N=10^3$ and $10^4$ at $T=\SI{2.0}{\second}$ to determine whether sufficient spatial margin is retained.

\paragraph{Maximum Ermakov Extent.}
We define the TF-equivalent extent along laboratory axis $i$ as
\begin{equation}
    E_i^{\rm GPE}(t)=\sqrt{7\Sigma_{ii}(t)},
\end{equation}
where $\Sigma(t)$ is given by Eq.~\ref{eq:gpe-covariance}. The factor $\sqrt{7}$ follows from the relation $R_{{\rm TF},k}=\sqrt{7}\sigma_{{\rm TF},k}$ for a Thomas--Fermi profile (Appendix~\ref{appendix:tf-equilibrium}). For $N=10^3$, the maximum Ermakov extent closely matches the maximum TF-equivalent extent along $x$ and $y$ but underestimates the $z$ extent by approximately $16\%$. For $N=10^4$, the agreement is closer, with a maximum discrepancy of $3.8\%$ across the three axes.

\paragraph{Initial GPE State Extent.}
At $t=0$, the initial GPE extent $E_i^{\rm GPE}(0)$ is evaluated from the covariance of the initial GPE state obtained by imaginary-time relaxation. For comparison, we define the laboratory-axis Thomas--Fermi extent at time $t$ as
\begin{equation}
    E_i^{\rm TF}(t)
    =
    \sqrt{
        \sum_{k=1}^{3}
        R_{{\rm TF},k}^{2}(t)
        \left[\hat{\mathbf e}_k(t)\right]_i^{2}
    },
    \qquad
    i\in\{x,y,z\},
    \label{eq:tf-lab-extent}
\end{equation}
where $R_{{\rm TF},k}(t)$ and $\hat{\mathbf e}_k(t)$ are the Thomas--Fermi radius and trap principal axis at time $t$, respectively. The initial value $E_i^{\rm TF}(0)$ provides the extent used to initialize the Ermakov scaling model (Appendix~\ref{appendix:tf-scaling}).

Table~\ref{tab:initial-extent-ratios} shows that the initial GPE $z$ extent exceeds $E_z^{\rm TF}(0)$ by $13.8\%$ at $N=10^3$ and $3.1\%$ at $N=10^4$. Thus, most of the later $z$-axis underestimation is already present in the initial scaling-model estimate rather than generated during transport.
\begin{table}[!htbp]
    \centering
    \small
    \setlength{\tabcolsep}{6pt}
    \begin{tabular}{cccc}
        \toprule
        & \multicolumn{3}{c}{$E_i^{\rm GPE}(0)/E_i^{\rm TF}(0)$} \\
        \cmidrule(lr){2-4}
        $N$ & $x$ & $y$ & $z$ \\
        \midrule
        $10^3$ & 0.957 & 1.014 & 1.138 \\
        $10^4$ & 0.987 & 1.000 & 1.031 \\
        \bottomrule
    \end{tabular}
    \caption{Initial GPE/Thomas--Fermi extent ratios for the selected $T=\SI{2.0}{\second}$ cases.}
    \label{tab:initial-extent-ratios}
\end{table}

\paragraph{Additional Budgets.}
The tail margin $M_{\rm tail}$ accounts for most of the additional transverse budget seen in Fig.~\ref{fig:sizer-adequacy}. Compared with $N=10^3$, the $N=10^4$ case has a smaller $M_{\rm tail}$ because of its smaller healing length under the heuristic tail rule (Eq.~\ref{eq:tail-margin}). This rule is intended to provide a boundary margin rather than a quantitative model of the GPE tail, so the decrease of $M_{\rm tail}$ with $N$ should not be interpreted as an established physical scaling. Nevertheless, the margin is important for the evaluated cases because omitting it would make the $z$-axis budget smaller than the maximum cloud-plus-COM extent at both atom numbers.

The same inverse-optimized trap trajectory is used for both atom numbers, so their node-to-node trap displacements and hence their $D_i^{\rm shift}$ contributions are identical. The Gaussian COM dynamics are independent of $N$, so their $S_i^{\rm COM}$ contributions are also identical. Along $x$, the translation buffer $D_x^{\rm shift}=\SI{1.255}{\micro\meter}$ is another substantial contribution to the grid-sizing budget. Although this buffer may appear generous, achieving a smaller value for the same trap trajectory would require a control interval below $\Delta t_{\rm ctrl}=\SI{0.5}{\milli\second}$, increasing the computational cost of the inverse optimization.

Although the larger cloud occupies a greater fraction of the grid-sizing budget at $N=10^4$, the surrogate-assisted grid-sizing construction provides spatial margin in both cases.

\subsection{Spatial and Spectral Occupancy}

We monitor the propagated states using spatial and spectral occupancy diagnostics.

\begin{figure}[!htbp]
    \centering
    \includegraphics[width=0.8\textwidth]{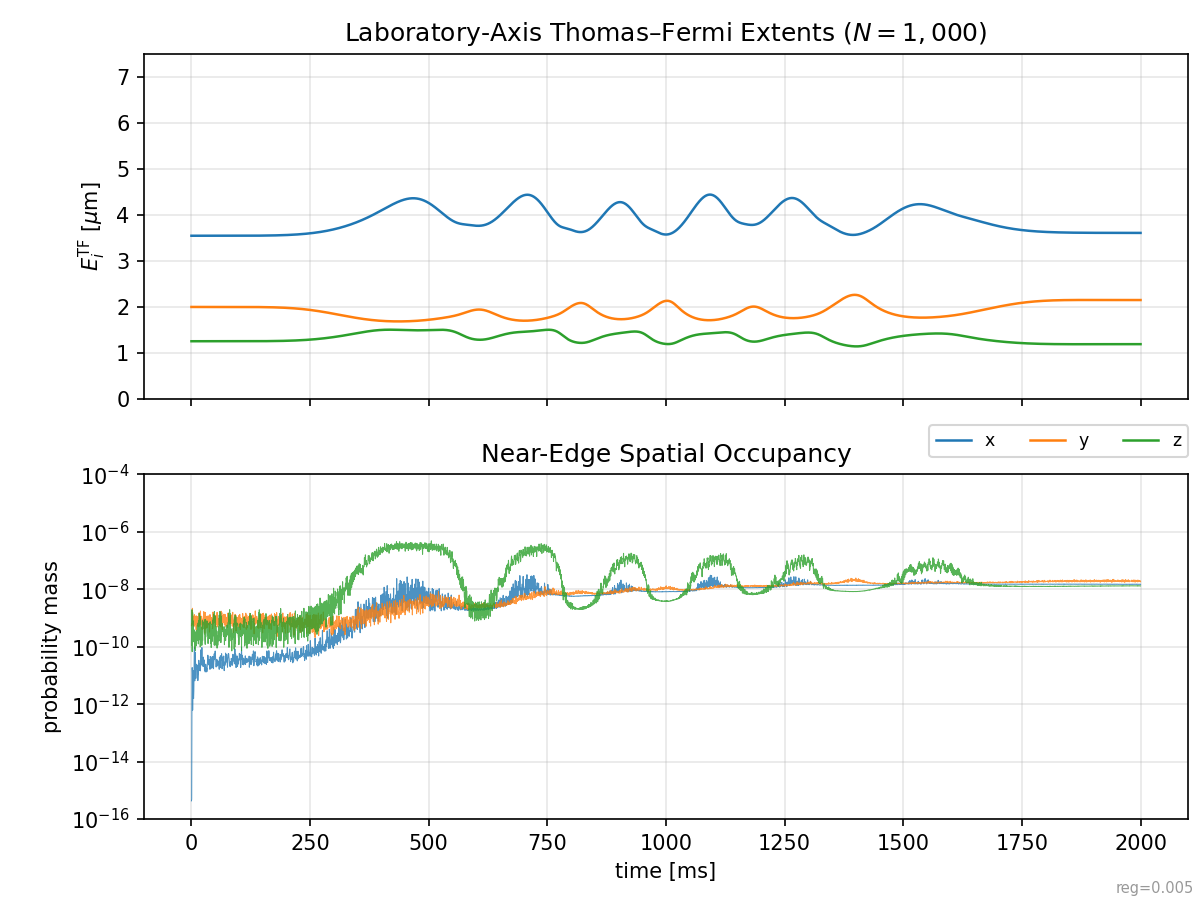}
    \caption{Laboratory-axis Thomas--Fermi extents and near-edge spatial occupancy for $N=10^3$ and $T=\SI{2.0}{\second}$.}
    \label{fig:spatial-occupancy-10e3}
\end{figure}

\paragraph{Spatial Occupancy.}
We define the near-edge spatial occupancy as the probability mass in the two outer regions $|r_i|\geq0.85L_i/2$ along each grid direction. Figure~\ref{fig:spatial-occupancy-10e3} compares this occupancy with the laboratory-axis Thomas--Fermi extents (Eq.~\ref{eq:tf-lab-extent}) for $N=10^3$ and $T=\SI{2.0}{\second}$.
The repeated extent variations reflect changes in confinement as the trap moves across the discrete shifting wires. The $z$-axis occupancy follows the same sequence of peaks as $E_z^{\rm TF}$, supporting reversible expansion driven by the changing confinement rather than accumulating boundary contamination. 

\begin{figure}[!htbp]
    \centering
    \includegraphics[width=0.8\textwidth]{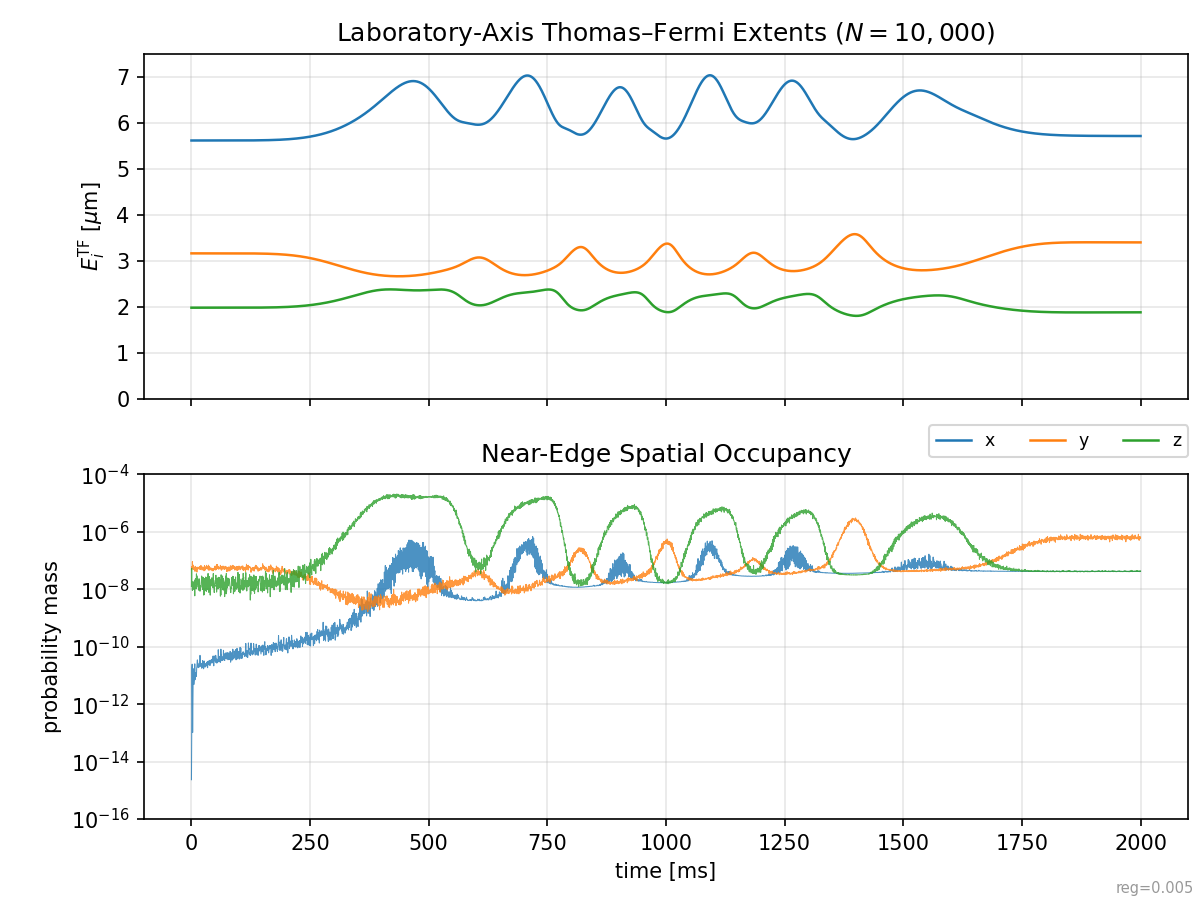}
    \caption{Laboratory-axis Thomas--Fermi extents and near-edge spatial occupancy for $N=10^4$ and $T=\SI{2.0}{\second}$.}
    \label{fig:spatial-occupancy-10e4}
\end{figure}

The predominance of the $z$ contribution is also consistent with the grid-sizing analysis above, where the GPE $z$-extent exceeds the surrogate estimate and the $z$-axis sizing budget relies most strongly on the tail margin. Fig.~\ref{fig:spatial-occupancy-10e4} shows a similar temporal pattern at $N=10^4$, but the $z$-axis occupancy reaches $2.1\times10^{-5}$ before returning to a much smaller value at delivery. 

\begin{figure}[!htbp]
    \centering
    \begin{subfigure}{0.9\textwidth}
        \centering
        \includegraphics[width=\textwidth]{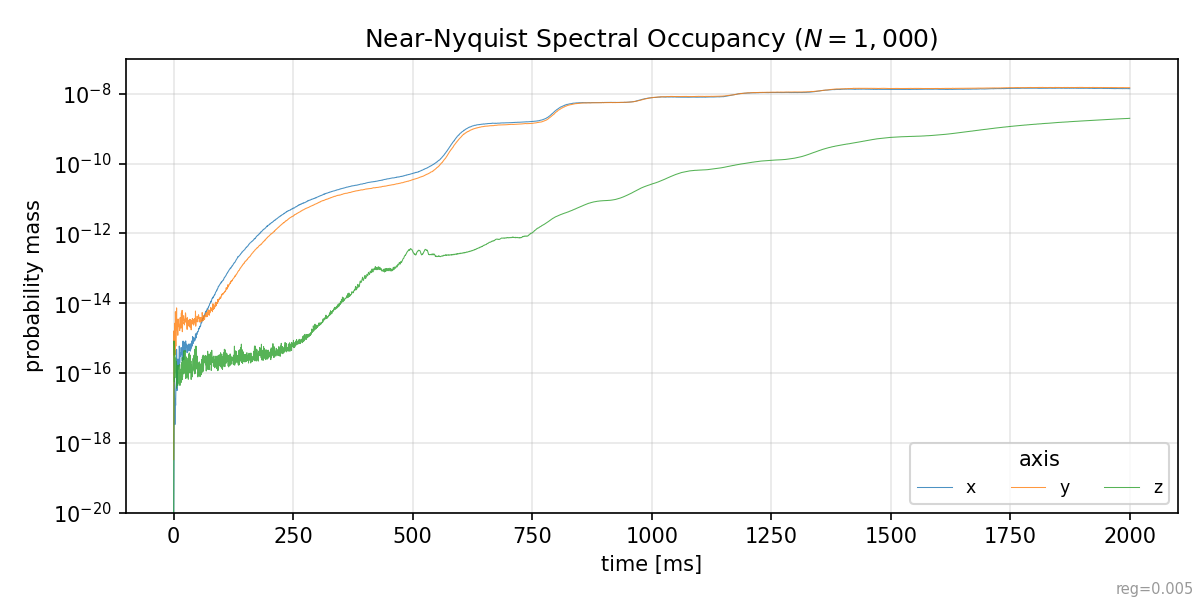}
    \end{subfigure}

    \smallskip

    \begin{subfigure}{0.9\textwidth}
        \centering
        \includegraphics[width=\textwidth]{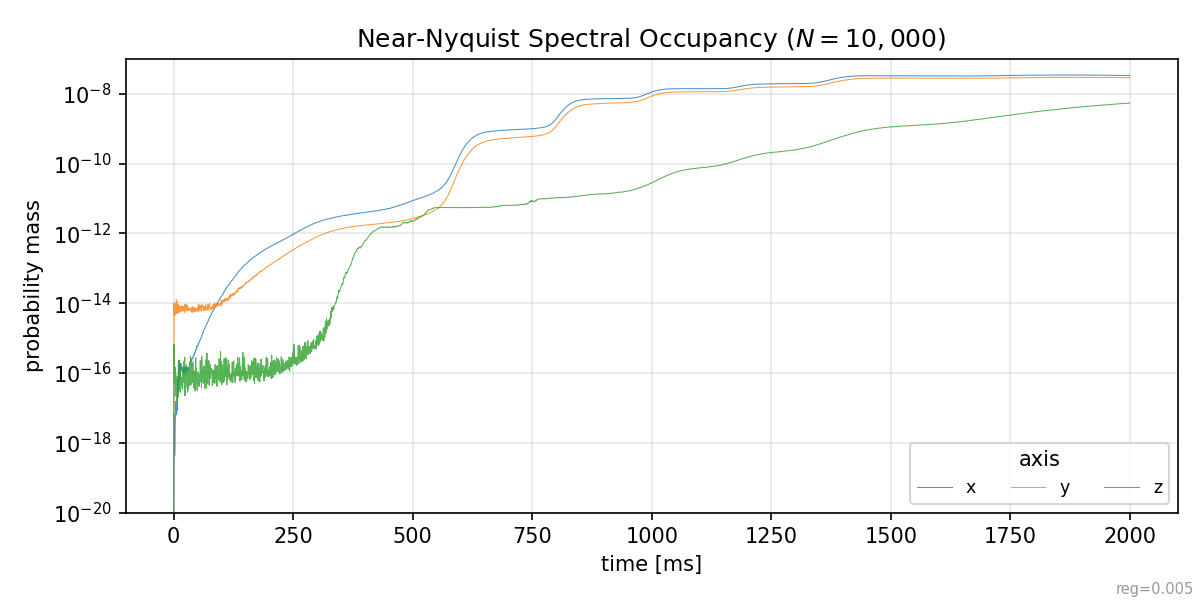}
    \end{subfigure}
    \caption{Near-Nyquist spectral occupancy for the selected $T=\SI{2.0}{\second}$ cases.}
    \label{fig:spectral-occupancy}
\end{figure}

\paragraph{Spectral Occupancy.}
We define the near-Nyquist spectral occupancy as the probability mass with $|k_i|\geq0.85k_{{\rm Nyq},i}$. Since wavevector components beyond the Nyquist limit would be aliased into the represented range, this diagnostic indicates whether the momentum distribution approaches the spectral cutoff. Fig.~\ref{fig:spectral-occupancy} compares the near-Nyquist spectral occupancy for the $N=10^3$ and $N=10^4$ cases.
Although the near-Nyquist occupancy increases gradually during transport, it remains negligible.

\subsection{Spatial and Temporal Refinement}

To assess whether the spatial and temporal discretizations materially affect the reported endpoint fidelity, we perform a combined refinement test for the $N=10^3$, $T=\SI{2.0}{\second}$ case. The baseline calculation uses $\Delta r_i\leq\xi_{\min}/2$ (Eq.~\ref{eq:spatial-resolution}) and $\Delta t=\SI{0.5}{\micro\second}$ (Table~\ref{tab:phase-rates}). Keeping the physical box, schedule, and relaxation rule fixed, we tighten the target spacing to $\xi_{\min}/3$ and the time step to $\SI{0.2}{\micro\second}$.

The grid therefore increases from $125\times70\times56$ to $189\times100\times84$ cells, while $n_{\rm sub}$ increases from 1000 to 2500, raising the nominal grid-point--step workload by a factor of approximately $8.1$. Despite this additional cost, the refined calculation increases $F_{3D}$ by only $5.2\times10^{-8}$.

\section{Conclusion}

We have presented a surrogate-assisted workflow for characterizing Bose--Einstein condensate transport on an atom chip. Regularized inverse optimization generates candidate wire-current schedules, a Gaussian phase-space model ranks them by estimated endpoint fidelity, and Gaussian and Thomas--Fermi scaling estimates provide case-specific grid dimensions before selected schedules are evaluated with the three-dimensional GPE.

At $N=10^3$, the surrogate and GPE distinguish the appreciable endpoint excitation of the $T=\SI{0.5}{\second}$ transport from the near-unit delivery of the longer transports, for which every selected schedule gives $F_{3D}>0.9997$. Their maximum axial COM excursions differ by at most \SI{5.6}{\nano\meter}, supporting the quadratic approximation used for surrogate COM propagation. Nevertheless, the COM surrogate and linear calculation generally predict much smaller endpoint losses than the GPE for the longer transports. The GPE width dynamics also show larger variations and superimposed oscillations absent from the linear calculation, demonstrating the importance of interaction-dependent internal dynamics beyond COM motion.

Because the $N=10^4$ calculation is substantially more computationally demanding, we evaluate it only at $T=\SI{2.0}{\second}$, which gives the highest GPE endpoint fidelity among the selected $N=10^3$ schedules. This case achieves near-unit fidelity, $F_{3D}=0.999932$, and has similar relative width ranges despite its larger absolute cloud widths.

The numerical-adequacy analysis shows that the combined surrogate-assisted grid-sizing budget retains spatial margin at both atom numbers, including along $z$ despite the Ermakov model's underestimation of the maximum GPE extent. The near-Nyquist spectral occupancy remains negligible, and, at $N=10^3$, a combined spatial and temporal refinement increases the nominal grid-point--step workload by a factor of approximately $8.1$ while changing $F_{3D}$ by only $5.2\times10^{-8}$.

Future work could extend surrogate screening to candidate schedules based on other trajectory-design methods, such as shortcuts to adiabaticity~\cite{mishra2025shortcuts}, since the screening stage does not require inverse optimization. Moreover, incorporating predicted density and flow dynamics could extend the endpoint-fidelity estimate beyond the COM-only description. Another direction is to include three-body recombination~\cite{soding1999threebody} in the GPE, allowing atom loss and its effect on the interaction strength to be assessed, particularly at higher densities and over longer transport durations. These extensions would bring the workflow closer to the full condensate dynamics while aiming to retain the computational advantage of surrogate screening.

\section*{Data Availability}
The code and data are available at \url{https://github.com/naokishibuya/atom-chip-optimizer-gpe}.

\section*{Acknowledgments}

The author thanks Dr Fedja Oručević and the Quantum Systems and Devices laboratory at the University of Sussex for supervision of the master’s thesis~\cite{shibuya2025arxiv} that this work extends and for providing the original MATLAB atom-chip trap model, from which the chip geometry and trap parameters used here are taken.

\clearpage
\appendix
\titleformat{\section}
  {\normalfont\Large\bfseries}
  {Appendix \thesection:}
  {0.5em}
  {}

\section{Gross--Pitaevskii Equation}\label{appendix:gpe-derivation}

\subsection{Many-Body Hamiltonian and Hartree Energy Functional}

We obtain the Hartree energy functional from the many-body Hamiltonian.

\paragraph{Many-Body Hamiltonian.}
The many-body Hamiltonian contains the kinetic energy of each atom, the time-dependent external potential $U(\mathbf r_i,t)$, and the pairwise interaction potential $U_{\rm int}(\mathbf r_i-\mathbf r_j)$,
\begin{equation}
    \hat H(t)
    =
    \sum_{i=1}^{N} \hat{h}(\mathbf r_i,t)
    +
    \sum_{i<j}
    U_{\rm int}(\mathbf r_i-\mathbf r_j),
    \qquad
    \hat h(\mathbf r,t)=-\frac{\hbar^2}{2m}\nabla^2+U(\mathbf r,t),
\end{equation}
where $N$ is atom number, $m$ is atom mass, and $\mathbf r_i$ is the position of atom $i$. $\sum_{i<j} U_{\rm int}(\mathbf r_i-\mathbf r_j)$ includes each interacting pair once. The exact many-body wavefunction $\Psi_{\rm exact}(\mathbf r_1,\ldots,\mathbf r_N,t)$ obeys
\begin{equation}
    i\hbar\frac{\partial\Psi_{\rm exact}}{\partial t}
    =
    \hat H(t)\Psi_{\rm exact}.
\end{equation}
Direct solution is impractical for the atom numbers considered here because $\Psi_{\rm exact}$ depends on all $3N$ spatial coordinates. For a dilute Bose--Einstein condensate with negligible thermal and interaction-induced depletion, the many-body wavefunction can instead be approximated by a Hartree product.

\paragraph{Hartree Energy Functional.}
The many-body wavefunction is written as a product of identical single-particle wavefunctions $\psi(\mathbf r,t)$,
\begin{equation}
    \Psi_{\rm H}(\mathbf r_1,\ldots,\mathbf r_N,t)
    =
    \prod_{i=1}^{N}\psi(\mathbf r_i,t), \qquad \int |\psi(\mathbf r, t)|^2\,d^3\mathbf r=1 .
\end{equation}
The total energy expectation value in the Hartree-product state can be decomposed as
\begin{equation}
    E_{\rm tot}[\psi,t]
    =
    \left\langle
        \Psi_{\rm H}
        \middle|
        \hat H(t)
        \middle|
        \Psi_{\rm H}
    \right\rangle \\[1ex]
    =
    E_{\rm one}[\psi,t]
    +
    E_{\rm int}[\psi,t],
\end{equation}
where $E_{\rm one}[\psi,t]$ is the one-body contribution
\begin{equation}
    E_{\rm one}[\psi,t]
    =
    \left\langle
        \Psi_{\rm H}
        \middle|
        \sum_{i=1}^{N} \hat h(\mathbf r_i, t)
        \middle|
        \Psi_{\rm H}
    \right\rangle
    =
    N\int
    \psi^*(\mathbf r,t)\hat h(\mathbf r, t)\psi(\mathbf r,t)
    \,d^3\mathbf r,
\end{equation}
and $E_{\rm int}[\psi, t]$ is the interaction contribution
\begin{equation}
    E_{\rm int}[\psi,t]
    =
    \left\langle
        \Psi_{\rm H}
        \middle|
        \sum_{i<j}
        U_{\rm int}(\mathbf r_i-\mathbf r_j)
        \middle|
        \Psi_{\rm H}
    \right\rangle
    =
    \frac{N(N-1)}{2}
    \iint
    U_{\rm int}(\mathbf r-\mathbf r')
    |\psi(\mathbf r,t)|^2
    |\psi(\mathbf r',t)|^2
    \,d^3\mathbf r\,d^3\mathbf r'.
\end{equation}
Averaging over the other particle coordinate defines the Hartree potential,
\begin{equation}
    U_{\rm H}[\psi](\mathbf r,t)
    =
    \int
    U_{\rm int}(\mathbf r-\mathbf r')
    |\psi(\mathbf r',t)|^2
    \,d^3\mathbf r',
    \label{eq:hartree-potential}
\end{equation}
which represents the averaged pair interaction.

Dividing the total energy $E_{\rm tot} = E_{\rm one}+E_{\rm int}$ by $N$ gives the energy functional per atom
\begin{equation}
    E[\psi,t]
    =
    \int
    \psi^*(\mathbf r,t)
    \left[
        \hat h(\mathbf r, t)
        +
        \frac{N-1}{2}
        U_{\rm H}[\psi](\mathbf r,t)
    \right]
    \psi(\mathbf r,t)
    \,d^3\mathbf r.
    \label{eq:hartree-energy-functional}
\end{equation}

\subsection{Variational Condition and Time-Dependent GPE}

We derive the time-dependent GPE by applying a variational principle to the Hartree energy functional.

\paragraph{Variational Condition.}
We define the residual of the many-body Schr\"odinger equation as
\begin{equation}
    R[\Psi]
    =
    \left(
        i\hbar\frac{\partial}{\partial t}
        -
        \hat H(t)
    \right)\Psi.
\end{equation}
The exact wavefunction satisfies $R[\Psi_{\rm exact}]=0$. To determine the closest evolution available in Hartree-product form, we introduce a variation $\delta\psi=\epsilon\eta$, where $\eta$ specifies the direction and $\epsilon$ its small magnitude.
\begin{equation}
    \begin{aligned}
        \Psi_{\rm H}[\psi+\delta\psi]
        &=
        \prod_{i=1}^{N}
        \left[
            \psi(\mathbf r_i,t)
            +
            \delta\psi(\mathbf r_i,t)
        \right]\\
        &=
        \Psi_{\rm H}[\psi]
        +
        \sum_{i=1}^{N}
        \delta\psi(\mathbf r_i,t)
        \prod_{j\ne i}\psi(\mathbf r_j,t)
        +
        \mathcal O(\epsilon^2).
    \end{aligned}
\end{equation}
The corresponding variation of $\Psi_{\rm H}$ is therefore
\begin{equation}
    \delta\Psi_{\rm H}(\mathbf r_1,\ldots,\mathbf r_N,t)
    =
    \sum_{i=1}^{N}
    \delta\psi(\mathbf r_i,t)
    \prod_{j\ne i}\psi(\mathbf r_j,t).
\end{equation}
If the residual had a component along any such variation, changing $\psi$ in that direction could reduce it further. Thus, we require the residual to be orthogonal to every such variation,
\begin{equation}
    \left\langle
    \delta\Psi_{\rm H}
    \middle|
    R[\Psi_{\rm H}]
    \right\rangle
    =0.
\end{equation}
Normalization of the varied wavefunction gives
\begin{equation}
    \begin{aligned}
    1 
    &=
    \langle
    \psi+\delta\psi
    |
    \psi+\delta\psi
    \rangle \\[1ex]
    &=
    \langle\psi|\psi\rangle
    +
    \langle\delta\psi|\psi\rangle
    +
    \langle\psi|\delta\psi\rangle
    +
    \mathcal O(\epsilon^2)\\[1ex]
    &=
    1
    +
    2\operatorname{Re}
    \langle\delta\psi|\psi\rangle
    +
    \mathcal O(\epsilon^2).
    \end{aligned}
\end{equation}
This requires $\operatorname{Re}\langle\delta\psi|\psi\rangle=0$. Decomposing the variation into components orthogonal and parallel to $\psi$,
\begin{equation}
    \delta\psi
    =
    \delta\psi_\perp
    +
    \delta\psi_\parallel,
    \qquad
    \langle\delta\psi_\perp|\psi\rangle=0,
    \qquad
    \delta\psi_\parallel=c\psi.
\end{equation}
The normalization condition then applies only to the parallel component,
\begin{equation}
    \langle\delta\psi|\psi\rangle
    =
    \langle\delta\psi_\parallel|\psi\rangle
    =
    c^*\langle\psi|\psi\rangle
    =
    c^*.
\end{equation}
Hence $\operatorname{Re}[c]=0$, so $c=i\alpha$ with $\alpha\in\mathbb R$ and $\alpha = \mathcal{O}(\epsilon)$. Considering the parallel variation alone gives
\begin{equation}
    \psi+\delta\psi_\parallel
    =
    (1+i\alpha)\psi
    =
    e^{i\alpha}\psi
    +
    \mathcal O(\epsilon^2),
\end{equation}
which represents an infinitesimal global-phase change. Because this change does not affect the physical state, we fix the phase freedom by setting $\alpha=0$ in the variational condition, leaving only the orthogonal component of the variation,
\begin{equation}
    \delta\psi=\delta\psi_\perp,
    \qquad
    \langle\delta\psi|\psi\rangle=0.
\end{equation}

\paragraph{Time-Dependent Hartree Equation.}
The two terms in the orthogonality condition are
\begin{equation}
    0
    =
    \left\langle
        \delta\Psi_{\rm H}
        \middle|
        R[\Psi_{\rm H}]
    \right\rangle
    =
    i\hbar
    \left\langle
        \delta\Psi_{\rm H}
        \middle|
        \frac{\partial\Psi_{\rm H}}{\partial t}
    \right\rangle
    -
    \left\langle
        \delta\Psi_{\rm H}
        \middle|
        \hat H(t)
        \middle|
        \Psi_{\rm H}
    \right\rangle.
\end{equation}
The time-derivative term further separates into two contributions,
\begin{equation}
    \begin{aligned}
        \left\langle
            \delta\Psi_{\rm H}
            \middle|
            \frac{\partial\Psi_{\rm H}}{\partial t}
        \right\rangle
        &=
        N
        \left\langle
            \delta\psi
            \middle|
            \frac{\partial\psi}{\partial t}
        \right\rangle
        \langle\psi|\psi\rangle^{N-1}
        \,+
        N(N-1)
        \langle\delta\psi|\psi\rangle
        \left\langle
            \psi
            \middle|
            \frac{\partial\psi}{\partial t}
        \right\rangle
        \langle\psi|\psi\rangle^{N-2}\\[1ex]
        &=
        N\int
        \delta\psi^*(\mathbf r,t)
        \frac{\partial\psi(\mathbf r,t)}{\partial t}
        \,d^3\mathbf r,
    \end{aligned}
\end{equation}
where the final equality uses $\langle\psi|\psi\rangle=1$ and $\langle\delta\psi|\psi\rangle=0$. The Hamiltonian also splits into two terms,
\begin{equation}
    \begin{aligned}
        \left\langle
            \delta\Psi_{\rm H}
            \middle|
            \hat H(t)
            \middle|
            \Psi_{\rm H}
        \right\rangle
        &=
        N\langle\delta\psi|\hat h(\mathbf r,t)|\psi\rangle
        +
        N(N-1)
        \langle\delta\psi|U_{\rm H}[\psi]|\psi\rangle\\
        &=
        N\int
        \delta\psi^*(\mathbf r,t)
        \left[
            \hat h(\mathbf r,t)
            +(N-1)U_{\rm H}[\psi](\mathbf r,t)
        \right]
        \psi(\mathbf r,t)
        \,d^3\mathbf r,
    \end{aligned}
\end{equation}
where the two variation positions cancel the pair-counting factor $1/2$. The functional derivative of the Hartree energy per atom $E[\psi,t]$ (Eq.~\ref{eq:hartree-energy-functional}) with respect to the complex-conjugate field $\psi^*(\mathbf r,t)$ is
\begin{equation}
    \frac{\delta E[\psi,t]}
         {\delta\psi^*(\mathbf r,t)}
    =
    \left[
        \hat h(\mathbf r,t)
        +(N-1)U_{\rm H}[\psi](\mathbf r,t)
    \right]\psi(\mathbf r,t).
\end{equation}
The orthogonality condition is then
\begin{equation}
    \int
    \delta\psi^*(\mathbf r,t)
    \left\{
        i\hbar\frac{\partial\psi(\mathbf r,t)}{\partial t}
        -
        \frac{\delta E[\psi,t]}
             {\delta\psi^*(\mathbf r,t)}
    \right\}
    \,d^3\mathbf r = 0.
\end{equation}
Since the allowed variations span the subspace orthogonal to $\psi$, the quantity in braces must be parallel to $\psi$, but need not be zero. Writing the parallel term as $\gamma(t)\psi$, with $\gamma(t)\in\mathbb C$, its imaginary part would cause the norm of $\psi$ to grow or decay. Norm conservation therefore requires $\gamma(t)$ to be real, leaving only a time-dependent global-phase change, which we absorb into $\psi$. The time evolution within the Hartree-product approximation is then
\begin{equation}
    i\hbar\frac{\partial\psi(\mathbf r,t)}{\partial t}
    =
    \frac{\delta E[\psi,t]}
         {\delta\psi^*(\mathbf r,t)}.
    \label{eq:energy-functional-time-derivative}
\end{equation}

\paragraph{Time-Dependent GPE.}
For a dilute ultracold gas, low-energy two-body collisions are dominated by $s$-wave scattering. Matching the contact pseudopotential to the scattering length $a_s$ and using the Hartree potential (Eq.~\ref{eq:hartree-potential}) gives
\begin{equation}
    U_{\rm int}(\mathbf r-\mathbf r')
    =
    g_0\delta(\mathbf r-\mathbf r'),
    \quad
    g_0
    =
    \frac{4\pi\hbar^2a_s}{m}
    \qquad\Longrightarrow\qquad
    U_{\rm H}[\psi](\mathbf r,t)
    =
    g_0|\psi(\mathbf r,t)|^2.
\end{equation}
With $g_{\rm eff}=(N-1)g_0$ ($\simeq Ng_0$ for $N\gg1$), Eq.~\ref{eq:hartree-energy-functional} becomes
\begin{equation}
    E[\psi,t]
    =
    \int
    \psi^*(\mathbf r,t)
    \left[
        \hat h(\mathbf r, t)
        +
        \frac{g_{\rm eff}}{2}|\psi(\mathbf r,t)|^2
    \right]
    \psi(\mathbf r,t)
    \,d^3\mathbf r.
    \label{eq:energy-functional-contact-interaction}
\end{equation}
Applying Eq.~\ref{eq:energy-functional-time-derivative} to this energy functional and expanding $\hat h(\mathbf r,t)$ gives the time-dependent GPE,
\begin{equation}
    i\hbar\frac{\partial\psi(\mathbf r,t)}{\partial t}
    =
    \left[
        -\frac{\hbar^2}{2m}\nabla^2
        +
        U(\mathbf r,t)
        +
        g_{\rm eff}|\psi(\mathbf r,t)|^2
    \right]\psi(\mathbf r,t).
    \label{eq:time-dependent-gpe}
\end{equation}

\subsection{Stationary GPE and Chemical Potential}

\paragraph{Stationary GPE.}
For a time-independent external potential $U(\mathbf r)$, the condensate ground state is obtained by minimizing $E[\psi]$ subject to normalization. Introducing a Lagrange multiplier $\mu$ gives the constrained functional
\begin{equation}
    \mathcal E_\mu[\psi]
    =
    E[\psi]
    -
    \mu\left(
        \int|\psi|^2\,d^3\mathbf r-1
    \right).
\end{equation}
Treating $\psi$ and $\psi^*$ as independent variables and setting
$\delta\mathcal E_\mu/\delta\psi^*=0$ gives
\begin{equation}
    0
    =
    \frac{\delta\mathcal E_\mu}{\delta\psi^*}
    =
    \frac{\delta E}{\delta\psi^*}
    -
    \mu\psi.
    \label{eq:stationary-variational-condition}
\end{equation}
Using Eqs.~\ref{eq:energy-functional-time-derivative} and \ref{eq:time-dependent-gpe} to evaluate the functional derivative gives the stationary GPE,
\begin{equation}
    \left[
        -\frac{\hbar^2}{2m}\nabla^2
        +
        U(\mathbf r)
        +
        g_{\rm eff}|\psi|^2
    \right]\psi
    =
    \mu\psi.
    \label{eq:stationary-gpe}
\end{equation}

\paragraph{Energy Variation with Atom Number.}
The total stationary energy with $N$ atoms is
\begin{equation}
    E_{\rm tot}[\psi; N]
    =
    N\,E[\psi]
    =
    \int
    \left[
        -N\,\frac{\hbar^2}{2m}\psi^*\nabla^2\psi
        +
        N\,U(\mathbf r)|\psi|^2
        +
        \frac{N(N-1)g_0}{2}|\psi|^4
    \right]
    d^3\mathbf r.
\end{equation}
Treating $N$ as continuous, the total derivative of the equilibrium energy is given by the chain rule,
\begin{equation}
    \frac{dE_{\rm tot}}{dN}
    =
    \frac{\partial E_{\rm tot}}{\partial N}
    +
    \int
    \left(
        \frac{\delta E_{\rm tot}}{\delta\psi}\frac{\partial\psi}{\partial N}
        +
        \frac{\delta E_{\rm tot}}{\delta\psi^*}\frac{\partial\psi^*}{\partial N}
    \right)
    d^3\mathbf r.
\end{equation}
The first term represents the explicit $N$-dependence of the energy functional, while the integral accounts for the readjustment of the wavefunction as $N$ changes. Because $\psi(\mathbf r)$ is a field, the latter contribution uses functional derivatives integrated over space.

\paragraph{Chemical Potential.}
For a fixed number of atoms $N$, Eq.~\ref{eq:stationary-variational-condition} and its complex conjugate give
\begin{equation}
    \frac{\delta E_{\rm tot}}{\delta\psi}
    =
    N\mu\psi^*,
    \qquad
    \frac{\delta E_{\rm tot}}{\delta\psi^*}
    =
    N\mu\psi.
\end{equation}
Therefore, the wavefunction-dependent part of the chain rule becomes
\begin{equation}
    N\mu
    \int
    \left(
        \psi^* \frac{\partial \psi}{\partial N}
    +
        \psi \frac{\partial \psi^*}{\partial N}
    \right)
    d^3\mathbf r
    =
    N\mu
    \frac{d}{dN}
    \int|\psi|^2\,d^3\mathbf r
    =
    0,
\end{equation}
because the normalization remains unity. The energy change is thus given by the partial derivative,
\begin{equation}
    \frac{d E_{\rm tot}}{d N}
    =
    \frac{\partial E_{\rm tot}}{\partial N}
    =
    \int
    \left[
        -\frac{\hbar^2}{2m}\psi^*\nabla^2\psi
        +
        U(\mathbf r)|\psi|^2
        +
        \left(
            N - \frac{1}{2}
        \right)
        g_0
        \,|\psi|^4
    \right]
    d^3\mathbf r.
\end{equation}
Multiplying the stationary GPE (Eq.~\ref{eq:stationary-gpe}) by $\psi^*$ and integrating gives an expression for $\mu$.
\begin{equation}
    \int \psi^* \left[
    -\frac{\hbar^2}{2m}\nabla^2
    +
    U(\mathbf{r})
    +
    g_{\rm eff}|\psi|^2
    \right]\psi
    d^3\mathbf r
    =
    \mu.
\end{equation}
The interaction coefficients differ by $g_0/2$, which is negligible for $N \gg 1$, yielding the chemical potential
\begin{equation}
    \mu\simeq\frac{d E_{\rm tot}}{d N},
\end{equation}
which approximates the equilibrium total-energy change when one atom is added to the condensate.

\section{Numerical GPE Propagation}

\subsection{Strang Factorization}\label{appendix:operator-splitting}

\paragraph{Operator Splitting.}
The kinetic and effective-potential parts of the GPE are evolved separately.
\begin{equation}
    \hat T
    =
    -\frac{\hbar^2}{2m}\nabla^2,
    \qquad
    \hat V
    =
    U(\mathbf r,t)
    +
    g_{\rm eff}|\psi|^2.
\end{equation}
During the kinetic substep, only the kinetic term is retained,
\begin{equation}
    i\hbar\frac{\partial\psi(\mathbf r,t)}{\partial t}
    =
    \hat T\psi(\mathbf r,t),
    \qquad
    \psi(\mathbf r,t+\Delta t)
    =
    \exp\!\left(
        -i\frac{\hat T}{\hbar}\Delta t
    \right)\psi(\mathbf r,t).
\end{equation}
For a scalar $a$, the operator exponential is defined by its power series,
\begin{equation}
    \exp\!\left(
        a\hat T
    \right)\psi
    =
    \left(
        1
        +
        a\hat T
        +
        \frac{a^2}{2!}\hat T^2
        +
        \cdots
    \right)\psi.
\end{equation}
The Fourier transform $\mathcal F$ maps the wavefunction from the spatial to the wavevector representation,
\begin{equation}
    \widetilde\psi(\mathbf k,t)
    =
    \mathcal F[\psi]
    =
    \int
    e^{-i\mathbf k\cdot\mathbf r}
    \psi(\mathbf r,t)
    \,d^3\mathbf r.
\end{equation}
Each plane wave is an eigenfunction of $\hat T$,
\begin{equation}
    \hat T e^{i\mathbf k\cdot\mathbf r}
    =
    -\frac{\hbar^2}{2m}
    \nabla^2 e^{i\mathbf k\cdot\mathbf r}
    =
    E_{\mathbf k}e^{i\mathbf k\cdot\mathbf r},
    \qquad
    E_{\mathbf k}
    =
    \frac{\hbar^2|\mathbf k|^2}{2m}.
\end{equation}
Therefore, $\hat T^n e^{i\mathbf k\cdot\mathbf r} = E_{\mathbf k}^n e^{i\mathbf k\cdot\mathbf r}$. Using this relation in the power series of the kinetic propagator gives
\begin{equation}
    \exp\!\left(
        -i\frac{\hat T}{\hbar}\Delta t
    \right)
    e^{i\mathbf k\cdot\mathbf r}
    =
    \exp\!\left(
        -i\frac{E_{\mathbf k}}{\hbar}\Delta t
    \right)
    e^{i\mathbf k\cdot\mathbf r}.
\end{equation}
Thus, each Fourier component is updated pointwise as
\begin{equation}
    \widetilde\psi(\mathbf k,t+\Delta t)
    =
    \exp\!\left(
        -i\frac{\hbar|\mathbf k|^2}{2m}\Delta t
    \right)
    \widetilde\psi(\mathbf k,t).
\end{equation}
The inverse Fourier transform $\mathcal F^{-1}$ then returns the updated wavefunction to the spatial representation.

Conversely, $\hat V$ is diagonal in the spatial representation, so its propagator acts pointwise on $\psi(\mathbf r,t)$. During a potential stage, $U(\mathbf r,t)$ is held fixed and $\hat V$ is real. Its propagator changes only the local phase,
\begin{equation}
    \left|
        \exp\!\left(-i\frac{\hat V}{\hbar}\Delta t\right)
        \psi(\mathbf r,t)
    \right|^2
    =
    |\psi(\mathbf r,t)|^2.
\end{equation}
Consequently, evaluating $g_{\rm eff}|\psi|^2$ once at the beginning of the potential stage gives the exact update.

\paragraph{Splitting Error.}
To derive the splitting error for fixed linear operators, consider two generally noncommuting operators $\hat A$ and $\hat B$.
\begin{equation}
    e^{\Delta t(\hat A+\hat B)}
    =
    1
    +
    \Delta t(\hat A+\hat B)
    +
    \frac{\Delta t^2}{2}
    \left(
        \hat A^2
        +
        \hat A\hat B
        +
        \hat B\hat A
        +
        \hat B^2
    \right)
    +
    \mathcal O(\Delta t^3).
\end{equation}
A sequential split instead gives
\begin{equation}
    \begin{aligned}
        e^{\Delta t\hat A}e^{\Delta t\hat B}
        &=
        \left(
            1+\Delta t\hat A+\frac{\Delta t^2}{2}\hat A^2
            +
            \mathcal O(\Delta t^3)
        \right)
        \left(
            1+\Delta t\hat B+\frac{\Delta t^2}{2}\hat B^2
            +
            \mathcal O(\Delta t^3)
        \right)
        \\
        &=
        1
        +
        \Delta t(\hat A+\hat B)
        +
        \frac{\Delta t^2}{2}\hat A^2
        +
        \Delta t^2\hat A\hat B
        +
        \frac{\Delta t^2}{2}\hat B^2
        +
        \mathcal O(\Delta t^3).
    \end{aligned}
\end{equation}
The difference is
\begin{equation}
    e^{\Delta t\hat A}e^{\Delta t\hat B}
    -
    e^{\Delta t(\hat A+\hat B)}
    =
    \frac{\Delta t^2}{2}
    [\hat A,\hat B]
    +
    \mathcal O(\Delta t^3),
\end{equation}
so the sequential split has local error $\mathcal O(\Delta t^2)$ when $[\hat A,\hat B]\ne0$.

For fixed kinetic and potential operators, $\hat A=-\frac{i}{\hbar}\hat T$ and $\hat B=-\frac{i}{\hbar}\hat V$. Because $\hat T$ and $\hat V$ generally do not commute, this local error applies to their sequential propagation. Over multiple time steps, the local errors accumulate to a global error $\mathcal O(\Delta t)$, making the sequential split a first-order method.

\paragraph{Symmetric Composition.}
The Strang factorization used in Eq.~\ref{eq:strang-step} has the symmetric form
\begin{equation}
    e^{\Delta t\hat A/2}
    e^{\Delta t\hat B}
    e^{\Delta t\hat A/2}.
\end{equation}
Expanding the individual factors gives
\begin{equation}
    \begin{aligned}
        e^{\Delta t\hat A/2}
        &=
        1
        +
        \frac{\Delta t}{2}\hat A
        +
        \frac{\Delta t^2}{8}\hat A^2
        +
        \mathcal O(\Delta t^3),
        \\[1ex]
        e^{\Delta t\hat B}
        &=
        1
        +
        \Delta t\hat B
        +
        \frac{\Delta t^2}{2}\hat B^2
        +
        \mathcal O(\Delta t^3).
    \end{aligned}
\end{equation}
Multiplying the first two factors gives
\begin{equation}
    \begin{aligned}
        e^{\Delta t\hat A/2}e^{\Delta t\hat B}
        &=
        1
        +
        \Delta t
        \left(
            \frac{\hat A}{2}
            +
            \hat B
        \right)\\
        &\qquad+
        \Delta t^2
        \left(
            \frac{\hat A^2}{8}
            +
            \frac{\hat A\hat B}{2}
            +
            \frac{\hat B^2}{2}
        \right)
        +
        \mathcal O(\Delta t^3).
    \end{aligned}
\end{equation}
Multiplying by the remaining kinetic half-step then gives
\begin{equation}
    e^{\Delta t\hat A/2}
    e^{\Delta t\hat B}
    e^{\Delta t\hat A/2}
    =
    1
    +
    \Delta t(\hat A+\hat B)
    +
    \frac{\Delta t^2}{2}
    \left(
        \hat A^2
        +
        \hat A\hat B
        +
        \hat B\hat A
        +
        \hat B^2
    \right)
    +
    \mathcal O(\Delta t^3).
\end{equation}

Comparing this result with the exact expansion above shows that the symmetric composition has local error $\mathcal O(\Delta t^3)$ and global error $\mathcal O(\Delta t^2)$ for fixed linear operators. The same orders apply to the symmetric composition of the exact kinetic and nonlinear-potential GPE subflows~\cite{javanainen2006splitstep}.

Using the node-centered holds (Eq.~\ref{eq:node-centered-current}) and the time-step relation (Eq.~\ref{eq:propagation-time-step}), each propagation block spans $\Delta t_{\rm ctrl}/2$ and contains $n_{\rm sub}/2$ Strang steps with the applied current held fixed. Within a block, adjacent kinetic half-steps combine into full kinetic steps, leaving half-steps only at the beginning and end. The resulting split-step sequence is therefore
\begin{equation*}
    \begin{aligned}
        \psi_{\rm in}
        &\longrightarrow
        \mathcal F
        \longrightarrow
        \text{kinetic half-step}
        \longrightarrow
        \mathcal F^{-1}
        \longrightarrow
        \text{potential step}\\
        &\longrightarrow
        \mathcal F
        \longrightarrow
        \text{kinetic full-step}
        \longrightarrow
        \mathcal F^{-1}
        \longrightarrow
        \text{potential step}
        \longrightarrow
        \cdots\\
        &\longrightarrow
        \mathcal F
        \longrightarrow
        \text{kinetic half-step}
        \longrightarrow
        \mathcal F^{-1}
        \longrightarrow
        \psi_{\rm out}.
    \end{aligned}
\end{equation*}

\subsection{Imaginary-Time Relaxation}\label{appendix:imaginary-time}

The initial and target ground states are obtained separately by holding the external potential fixed at the corresponding transport endpoint and applying imaginary-time relaxation. This is a numerical procedure for finding the ground state, not a physical evolution.

\paragraph{Linear Ground-State Filtering.}
Although the GPE is nonlinear, the principle of imaginary-time relaxation is most easily seen for a time-independent linear Hamiltonian. Let its orthonormal eigenstates $\phi_n$ have eigenvalues $E_n$, ordered so that $E_0$ is the ground-state energy,
\begin{equation}
    \hat H\phi_n
    =
    E_n\phi_n.
\end{equation}
In real time, each eigenstate evolves according to
\begin{equation}
    e^{-i\hat Ht/\hbar}\phi_n
    =
    e^{-iE_nt/\hbar}\phi_n.
\end{equation}
The exponential factor has unit magnitude and therefore changes only the phase of each eigenstate. Replacing $t$ by $-i\tau$ instead gives
\begin{equation}
    e^{-\hat H\tau/\hbar}\phi_n
    =
    e^{-E_n\tau/\hbar}\phi_n,
\end{equation}
where $\tau$ is a numerical relaxation coordinate distinct from the physical transport time $t$. Let the initial guess have the eigenstate expansion
\begin{equation}
    \psi_{\rm guess}
    =
    \sum_n c_n\phi_n,
    \qquad
    c_0\ne0,
\end{equation}
where $c_n=\langle\phi_n|\psi_{\rm guess}\rangle$ are generally complex coefficients. The condition $c_0\ne0$ ensures that the initial guess contains a ground-state component. Applying the imaginary-time propagator $e^{-\hat H\tau/\hbar}$ to this initial guess gives
\begin{equation}
    \begin{aligned}
        \psi(\mathbf r,\tau)
        &=
        \left[
            e^{-\hat H\tau/\hbar}\psi_{\rm guess}
        \right](\mathbf r)\\[1ex]
        &=
        \sum_n
        c_n e^{-E_n\tau/\hbar}
        \phi_n(\mathbf r)\\
        &=
        c_0e^{-E_0\tau/\hbar}
        \left[
            \phi_0(\mathbf r)
            +
            \sum_{n>0}
            \frac{c_n}{c_0}
            e^{-(E_n-E_0)\tau/\hbar}
            \phi_n(\mathbf r)
        \right].
    \end{aligned}
\end{equation}
The propagated wavefunction is normalized according to
\begin{equation}
    \psi(\mathbf r,\tau)
    \longmapsto
    \frac{
        \psi(\mathbf r,\tau)
    }{
        \sqrt{
            \int
            |\psi(\mathbf r',\tau)|^2
            \,d^3\mathbf r'
        }
    }.
\end{equation}
Applied to the propagated state above, this removes the magnitude of the prefactor $c_0e^{-E_0\tau/\hbar}$ while retaining its phase $c_0/|c_0|$. Since $E_n-E_0>0$ for $n>0$, every excited-state component vanishes as $\tau\to\infty$, and the normalized state approaches $(c_0/|c_0|)\phi_0$.

\paragraph{Nonlinear GPE Relaxation.}
For the GPE, the effective Hamiltonian changes with the evolving density $|\psi|^2$, so there is no fixed eigenbasis whose components acquire the independent exponential factors used in the linear argument above. We therefore perform the relaxation iteratively.

At each imaginary-time iteration, we apply a kinetic half-step, evaluate the nonlinear potential from the updated state, hold it fixed during the potential step, apply the second kinetic half-step, and normalize the resulting wavefunction. Repeating this procedure relaxes the initial guess toward the GPE ground state used in the simulations.

\section{Gaussian Phase-Space Model}
\label{appendix:gaussian-propagation}

We derive the harmonic COM propagation and Gaussian endpoint-fidelity estimate.

\paragraph{COM Motion Under Harmonic Confinement.}
The laboratory-frame cloud centroid is
\begin{equation}
    \bar{\mathbf r}(t)
    =
    \langle\hat{\mathbf r}\rangle
    =
    \left\langle
        \psi
        \middle|
        \hat{\mathbf r}
        \middle|
        \psi
    \right\rangle
    =
    \int
    \psi^*(\mathbf r,t)\,
    \mathbf r\,
    \psi(\mathbf r,t)
    \,d^3\mathbf r.
\end{equation}
Let $\hat T=\hat{\mathbf p}^{\,2}/(2m)$ and $\hat V=U+g_{\rm eff}|\psi|^2$ denote the kinetic and effective-potential terms in the GPE (Eq.~\ref{eq:gpe}). Differentiating the centroid then gives the first Ehrenfest relation,
\begin{equation}
    \begin{aligned}
        \dot{\bar{\mathbf r}}
        =
        \frac{d}{dt}\langle\hat{\mathbf r}\rangle
        &=
        \left\langle
            \frac{\partial\psi}{\partial t}
            \middle|
            \hat{\mathbf r}
            \middle|
            \psi
        \right\rangle
        +
        \left\langle
            \psi
            \middle|
            \hat{\mathbf r}
            \middle|
            \frac{\partial\psi}{\partial t}
        \right\rangle\\
        &=
        \frac{1}{i\hbar}
        \left\langle
            \left[
                \hat{\mathbf r},
                \hat T+\hat V
            \right]
        \right\rangle
        =
        \frac{1}{i\hbar}
        \left\langle
            \left[
                \hat{\mathbf r},
                \frac{\hat{\mathbf p}^{\,2}}{2m}
            \right]
        \right\rangle\\
        &=
        \frac{\langle\hat{\mathbf p}\rangle}{m}
        =
        \frac{\bar{\mathbf p}}{m},
    \end{aligned}
    \vspace{-3pt}
\end{equation}
where $\left[\hat{\mathbf r},\hat{\mathbf p}^{\,2}\right]=2i\hbar\hat{\mathbf p}$ and the position-dependent operator $\hat V$ commutes with $\hat{\mathbf r}$. Thus, the interaction term does not enter this relation and the centroid velocity equals the mean momentum divided by $m$.

Using $\bar{\mathbf p}=m\dot{\bar{\mathbf r}}$, differentiating the mean momentum gives the second Ehrenfest relation,
\begin{equation}
    \begin{aligned}
        m\ddot{\bar{\mathbf r}}
        =
        \frac{d}{dt}\langle\hat{\mathbf p}\rangle
        &=
        \left\langle
            \frac{\partial\psi}{\partial t}
            \middle|
            \hat{\mathbf p}
            \middle|
            \psi
        \right\rangle
        +
        \left\langle
            \psi
            \middle|
            \hat{\mathbf p}
            \middle|
            \frac{\partial\psi}{\partial t}
        \right\rangle\\
        &=
        \frac{1}{i\hbar}
        \left\langle
            \left[
                \hat{\mathbf p},
                \hat T+\hat V
            \right]
        \right\rangle
        =
        \frac{1}{i\hbar}
        \left\langle
            \left[
                \hat{\mathbf p},
                \hat{V}
            \right]
        \right\rangle\\
        &=
        -\langle
            \nabla\hat{V}
        \rangle
        =
        -\int
        \psi^*
        \left[
            \nabla U
            +
            g_{\rm eff}
            \nabla|\psi|^2
        \right]
        \psi
        \,d^3\mathbf r,
    \end{aligned}
    \vspace{-3pt}
\end{equation}
where we used $[\hat{\mathbf p},f]\psi = -i\hbar\nabla(f\psi) + i\hbar f\nabla\psi = -i\hbar(\nabla f)\psi$ for any position-dependent function $f(\mathbf r)$. The integral multiplying $g_{\rm eff}$ can be written as a surface term,
\begin{equation}
    \int
    |\psi|^2\nabla|\psi|^2
    \,d^3\mathbf r
    =
    \frac{1}{2}
    \int
    \nabla|\psi|^4
    \,d^3\mathbf r
    =
    \frac{1}{2}
    \oint_{S_\infty}
    |\psi|^4
    \,d\mathbf S
    =
    0,
\end{equation}
where the density of a localized trapped state vanishes at spatial infinity. Thus, the contact interaction contributes no net force, and the rate of change of the mean momentum equals the mean external force,
\begin{equation}
    m\ddot{\bar{\mathbf r}}
    =
    -\int
    |\psi|^2\nabla U
    \,d^3\mathbf r
    =
    -\langle\nabla U\rangle.
\end{equation}
Under the harmonic approximation (Eq.~\ref{eq:harmonic-potential}), $\nabla U(\mathbf r,t) = H(t) \left[ \mathbf r-\mathbf r_{\min}(t) \right]$, yielding
\begin{equation}
    \begin{aligned}
        m\ddot{\bar{\mathbf r}}
        &=
        -H(t)
        \int
        \psi^*
        \left[
            \mathbf r-\mathbf r_{\min}(t)
        \right]
        \psi
        \,d^3\mathbf r \\
        &=
        -H(t)
        \left[
            \bar{\mathbf r}(t)-\mathbf r_{\min}(t)
        \right].
    \end{aligned}
\end{equation}
Thus, the centroid equation of motion under harmonic confinement is unaffected by the contact interaction and independent of the cloud's internal dynamics~\cite{wu2014dynamics}. The relative COM motion therefore obeys
\begin{equation}
    \ddot{\mathbf r}_{\rm com}(t)
    =
    -K(t)\mathbf r_{\rm com}(t)
    -
    \ddot{\mathbf r}_{\min}(t),
    \qquad
    K(t)=\frac{H(t)}{m},
\end{equation}
where $\mathbf r_{\rm com}(t)=\bar{\mathbf r}(t)-\mathbf r_{\min}(t)$. This COM propagation does not assume a particular cloud shape. It requires only that the harmonic approximation remain valid over the region occupied by the cloud.

\paragraph{Constant-Current COM Propagation.}
With the augmented phase-space vector 
\begin{equation}
    \mathbf y
    =
    \begin{pmatrix}
        \bar{\mathbf r}\\
        \bar{\mathbf p}\\
        1
    \end{pmatrix},
\end{equation}
the quadratic dynamics during the constant-current hold associated with control node $j$ are
\begin{equation}
    \dot{\mathbf y}
    =
    \mathcal A_j\mathbf y,
    \qquad
    \mathcal A_j =
    \begin{pmatrix}
        0_{3\times3} & \mathbb I_3/m & 0_{3\times1}\\
        -H(t_j) & 0_{3\times3} & H(t_j)\mathbf r_{\min}(t_j)\\
        0_{1\times3} & 0_{1\times3} & 0
    \end{pmatrix}.
\end{equation}
Because $\mathcal A_j$ is constant during a hold, the linear system has the solution
\begin{equation}
    \mathbf y(t+\Delta t_{\rm hold})
    =
    \sum_{n=0}^{\infty}
    \frac{(\Delta t_{\rm hold})^n}{n!}\mathbf y^{(n)}(t)\\
    =
    \sum_{n=0}^{\infty}
    \frac{
        (\mathcal A_j\Delta t_{\rm hold})^n
    }{n!}
    \mathbf y(t)
    =
    \exp\!\left(
        \mathcal A_j\Delta t_{\rm hold}
    \right)
    \mathbf y(t).
\end{equation}
The node-centered convention uses two half-holds between adjacent control nodes,
\begin{equation}
    \mathbf y(t_{j+1})
    =
    \exp\!\left(
        \mathcal A_{j+1}\frac{\Delta t_{\rm ctrl}}{2}
    \right)
    \exp\!\left(
        \mathcal A_j\frac{\Delta t_{\rm ctrl}}{2}
    \right)
    \mathbf y(t_j).
\end{equation}

\paragraph{Endpoint-Fidelity Estimate.}
We define $F_{\rm COM}$ as an endpoint-fidelity estimate based only on the predicted residual COM displacement and momentum. It assumes that the delivered cloud matches the target in all other respects, including cloud size, shape, orientation, and internal phase structure.

A one-dimensional Gaussian wavefunction with rms width $\sigma$, displacement $d$, and momentum $p$ is
\begin{equation}
    \phi_{d,p}(x)
    =
    \frac{1}{(2\pi\sigma^2)^{1/4}}
    \exp\!\left[
        -\frac{(x-d)^2}{4\sigma^2}
    \right]
    \exp\!\left(
        i\frac{p}{\hbar}x
    \right).
\end{equation}
The overlap between the target profile $\phi_{0,0}$ and the delivered profile $\phi_{d,p}$ is
\begin{equation}
    \begin{aligned}
        \langle\phi_{0,0}|\phi_{d,p}\rangle
        &=
        \int_{-\infty}^{\infty}
        \phi_{0,0}^*(x)\phi_{d,p}(x)
        \,dx\\
        &=
        \int_{-\infty}^{\infty}
        \left[
            \frac{1}{(2\pi\sigma^2)^{1/4}}
            \exp\!\left(
                -\frac{x^2}{4\sigma^2}
            \right)
        \right]
        \left[
            \frac{1}{(2\pi\sigma^2)^{1/4}}
            \exp\!\left(
                -\frac{(x-d)^2}{4\sigma^2}
            \right)
            \exp\!\left(
                \frac{ipx}{\hbar}
            \right)
        \right]
        \,dx\\
        &=
        \frac{1}{\sqrt{2\pi\sigma^2}}
        \int_{-\infty}^{\infty}
        \exp\!\left[
            -\frac{x^2+(x-d)^2}{4\sigma^2}
            +
            \frac{ipx}{\hbar}
        \right]
        \,dx\\
        &=
        \frac{1}{\sqrt{2\pi\sigma^2}}
        \exp\!\left[
            -\frac{d^2}{8\sigma^2}
            -\frac{\sigma^2p^2}{2\hbar^2}
            +\frac{idp}{2\hbar}
        \right]
        \int_{-\infty}^{\infty}
        \exp\!\left[
            -\frac{1}{2\sigma^2}
            \left(
                x-\frac{d}{2}
                -\frac{i\sigma^2p}{\hbar}
            \right)^2
        \right]
        dx\\
        &=
        \exp\!\left[
            -\frac{d^2}{8\sigma^2}
            -\frac{\sigma^2p^2}{2\hbar^2}
            +\frac{idp}{2\hbar}
        \right].
    \end{aligned}
\end{equation}
The last Gaussian integral evaluates to $\sqrt{2\pi\sigma^2}$. Taking the squared magnitude of the overlap gives
\begin{equation}
    \left|
        \langle\phi_{0,0}|\phi_{d,p}\rangle
    \right|^2
    =
    \exp\!\left[
        -\frac{d^2}{4\sigma^2}
        -
        \frac{\sigma^2p^2}{\hbar^2}
    \right].
\end{equation}
Applying this expression independently along the three principal axes, using the Thomas--Fermi rms widths (Appendix~\ref{appendix:tf-equilibrium}) as spatial scales, gives the endpoint-fidelity estimate $F_{\rm COM}$ in Eq.~\ref{eq:com-overlap}.

\section{Thomas--Fermi Equilibrium}
\label{appendix:tf-equilibrium}

We derive the equilibrium Thomas--Fermi description of a condensate under harmonic confinement.

\paragraph{Thomas--Fermi Density and Radii.}
For a fixed external potential, the condensate ground state satisfies the stationary GPE (Eq.~\ref{eq:stationary-gpe}). In the Thomas--Fermi limit, its kinetic term is small compared with the external-potential and interaction terms over most of the cloud and is therefore neglected. For the repulsive interactions considered here, $g_{\rm eff}$ increases with $N$, so the Thomas--Fermi approximation typically becomes more accurate at larger atom number.

Let $\mu_{\rm TF}$ be the approximate chemical potential. We solve the remaining stationary GPE for $n_{\rm TF}(\mathbf r)=|\psi(\mathbf r)|^2$ to obtain the Thomas--Fermi probability density
\begin{equation}
    n_{\rm TF}(\mathbf r)
    =
    \frac{1}{g_{\rm eff}}
    \max\!\left\{
        \mu_{\rm TF}-U(\mathbf r),
        0
    \right\}.
    \label{eq:tf-density}
\end{equation}
Taking the trap minimum as the zero of potential energy, the harmonic approximation (Eq.~\ref{eq:harmonic-potential}) gives
\begin{equation}
    n_{\rm TF}(\mathbf r,t)
    =
    \frac{1}{g_{\rm eff}}
    \max\!\left\{
        \mu_{\rm TF}(t)
        -
        \frac{1}{2}
        [\mathbf r-\mathbf r_{\min}(t)]^\top
        H(t)
        [\mathbf r-\mathbf r_{\min}(t)],
        0
    \right\}.
    \label{eq:tf-harmonic-density}
\end{equation}
The TF radius $R_{{\rm TF},k}(t)$ is the distance along principal direction $\hat{\mathbf e}_k(t)$ from the trap minimum to where the Thomas--Fermi density $n_{\rm TF}(\mathbf r,t)$ vanishes.
\begin{equation}
    \mu_{\rm TF}(t)
    -
    \frac{1}{2}
    m\omega_k^2(t)\,R_{{\rm TF},k}^2(t)
    =
    0
    \qquad
    \Longrightarrow
    \qquad
    R_{{\rm TF},k}(t)
    =
    \sqrt{
        \frac{2\mu_{\rm TF}(t)}
            {m\omega_k^2(t)}
    }.
\end{equation}
Thus, once $\mu_{\rm TF}(t)$ is known, the trap frequencies determine the three TF radii.

\paragraph{Thomas--Fermi Chemical Potential.}
To determine $\mu_{\rm TF}(t)$, we introduce the scaled coordinates
\begin{equation}
    u_k
    =
    \frac{
        \hat{\mathbf e}_k(t)\cdot
        [\mathbf r-\mathbf r_{\min}(t)]
    }{
        R_{{\rm TF},k}(t)
    }.
    \label{eq:tf-scaled-coords}
\end{equation}
With $\mathbf u=(u_1,u_2,u_3)^\top$, the density becomes
\begin{equation}
    n_{\rm TF}(\mathbf r,t)
    =
    \frac{\mu_{\rm TF}(t)}{g_{\rm eff}}
    \left(
        1-\rho^2
    \right),
    \qquad
    \rho = |\mathbf u|\leq1.
    \label{eq:tf-scaled-density}
\end{equation}
To evaluate the normalization integral, we express the volume element in these scaled coordinates. At fixed $t$, their spatial differentials are
\begin{equation}
    du_k
    =
    \frac{
        \hat{\mathbf e}_k(t)\cdot d\mathbf r
    }{
        R_{{\rm TF},k}(t)
    }.
\end{equation}
Writing $\mathbf u$ in spherical coordinates with an arbitrary polar axis, the infinitesimal volume element is
\begin{equation}
    \begin{aligned}
        d^3\mathbf r
        &=
        \prod_{k=1}^{3}
        \left[
            R_{{\rm TF},k}(t)\,du_k
        \right]\\
        &=
        \left[
            \prod_{k=1}^{3}R_{{\rm TF},k}(t)
        \right]
        d^3\mathbf u\\
        &=
        \left[
            \prod_{k=1}^{3}R_{{\rm TF},k}(t)
        \right]
        \rho^2\sin\theta
        \,d\rho\,d\theta\,d\phi.
    \end{aligned}
    \label{eq:tf-volume-element}
\end{equation}
We now impose the normalization condition,
\begin{equation}
    \begin{aligned}
        1
        =
        \int
        n_{\rm TF}(\mathbf r,t)
        \,d^3\mathbf r
        &=
        \frac{\mu_{\rm TF}(t)}{g_{\rm eff}}
        \left[
            \prod_{k=1}^{3}R_{{\rm TF},k}(t)
        \right]
        \int_0^1
        \int_0^\pi
        \int_0^{2\pi}
        \left(
            1-\rho^2
        \right)
        \rho^2\sin\theta
        \,d\phi\,d\theta\,d\rho\\
        &=
        \frac{\mu_{\rm TF}(t)}{g_{\rm eff}}
        \left[
            \prod_{k=1}^{3}R_{{\rm TF},k}(t)
        \right]
        \int_0^1
        4\pi
        \left(
            1-\rho^2
        \right)
        \rho^2
        \,d\rho\\
        &=
        \frac{8\pi}{15}
        \frac{\mu_{\rm TF}(t)}{g_{\rm eff}}
        \prod_{k=1}^{3}R_{{\rm TF},k}(t).
    \end{aligned}
    \label{eq:tf-normalization}
\end{equation}
Substituting the Thomas--Fermi radii gives
\begin{equation}
    \prod_{k=1}^{3}
    R_{{\rm TF},k}(t)
    =
    \sqrt{
        \frac{2\mu_{\rm TF}(t)}
             {m\omega_1^2(t)}
    }
    \sqrt{
        \frac{2\mu_{\rm TF}(t)}
             {m\omega_2^2(t)}
    }
    \sqrt{
        \frac{2\mu_{\rm TF}(t)}
             {m\omega_3^2(t)}
    }
    =
    \left[
    \frac{
        2\mu_{\rm TF}(t)
    }{
        m\bar\omega^2(t)
    }
    \right]^{3/2},
\end{equation}
where $\bar\omega(t)=\left[\omega_1(t)\omega_2(t)\omega_3(t)\right]^{1/3}$ is the geometric-mean trap frequency. Substituting $g_{\rm eff}\simeq4\pi\hbar^2Na_s/m$ into the normalization condition gives
\begin{equation}
    1
    \simeq
    \frac{8\pi}{15}
    \frac{\mu_{\rm TF}(t)}
         {4\pi\hbar^2Na_s/m}
    \left[
        \frac{
            2\mu_{\rm TF}(t)
        }{
            m\bar\omega^2(t)
        }
    \right]^{3/2}
    =
    \frac{a_{\rm ho}(t)}
         {15Na_s}
    \left[
        \frac{
            2\mu_{\rm TF}(t)
        }{
            \hbar\bar\omega(t)
        }
    \right]^{5/2},
\end{equation}
where we introduced the harmonic-oscillator length,
\begin{equation}
    a_{\rm ho}(t)
    =
    \sqrt{
        \frac{\hbar}{m\bar\omega(t)}
    }.
\end{equation}
The chemical potential is thus
\begin{equation}
    \mu_{\rm TF}(t)
    \simeq
    \frac{\hbar\bar\omega(t)}{2}
    \left[
        \frac{15Na_s}{a_{\rm ho}(t)}
    \right]^{2/5}.
\end{equation}

\paragraph{Thomas--Fermi RMS Widths.}
The Thomas--Fermi density is symmetric about the trap minimum, so the variance equals the mean squared displacement. For the variance along principal axis $k$, we take the $u_k$ direction as the polar axis, giving $u_k=\rho\cos\theta$.
\begin{equation}
    \begin{aligned}
        \sigma_{{\rm TF},k}^2(t)
        &=
        \int
        \left\{
            \hat{\mathbf e}_k(t)\cdot
            [\mathbf r-\mathbf r_{\min}(t)]
        \right\}^2
        n_{\rm TF}(\mathbf r,t)
        \,d^3\mathbf r\\
        &=
        \int_{|\mathbf u|\leq1}
        \left(
            R_{{\rm TF},k}(t)u_k
        \right)^2
        \frac{\mu_{\rm TF}(t)}{g_{\rm eff}}
        \left(
            1-\rho^2
        \right)
        \,d^3\mathbf r && (\text{Eqs.~\ref{eq:tf-scaled-coords},~\ref{eq:tf-scaled-density}})\\
        &=
        R_{{\rm TF},k}^2(t)
        \frac{\mu_{\rm TF}(t)}{g_{\rm eff}}
        \left[
            \prod_{j=1}^{3}R_{{\rm TF},j}(t)
        \right]
        \int_0^1
        \int_0^\pi
        \int_0^{2\pi}
        \left(
            1-\rho^2
        \right)
        \rho^2
        u_k^2\sin\theta
        \,d\phi
        \,d\theta
        \,d\rho && (\text{Eq.~\ref{eq:tf-volume-element}})\\
        &=
        R_{{\rm TF},k}^2(t)
        \frac{15}{8\pi}
        \int_0^1
        \int_0^\pi
        \int_0^{2\pi}
        \left(
            1-\rho^2
        \right)
        \rho^2
        (\rho\cos\theta)^2
        \sin\theta
        \,d\phi\,d\theta\,d\rho && (\text{Eq.~\ref{eq:tf-normalization}})\\
        &=
        R_{{\rm TF},k}^2(t)
        \frac{15}{8\pi}
        \left[
            \int_0^{2\pi}d\phi
        \right]
        \left[
            \int_0^\pi
            \cos^2\theta\sin\theta
            \,d\theta
        \right]
        \int_0^1
        \rho^4
        \left(
            1-\rho^2
        \right)
        \,d\rho\\
        &=
        \frac{R_{{\rm TF},k}^2(t)}{7}.
    \end{aligned}
\end{equation}
Therefore, the Thomas--Fermi rms width is
\begin{equation}
    \sigma_{{\rm TF},k}(t)
    =
    \frac{R_{{\rm TF},k}(t)}{\sqrt{7}}.
\end{equation}

\section{Healing Length}
\label{appendix:healing-length}

We derive the healing length as a characteristic scale for choosing the GPE grid spacing.

\paragraph{Local Energy Scales.}
Consider a small region where the external potential is approximately constant, $U(\mathbf r)\simeq U_0$. We choose a spatially constant reference wavefunction, which has uniform density, no flow, and a vanishing kinetic term. We also set its global (constant) phase to zero,
\begin{equation}
    \psi_{\rm ref}
    =
    \sqrt{n_{\rm ref}},
    \qquad
    \nabla^2\psi_{\rm ref}
    =
    0.
\end{equation}
The stationary GPE (Eq.~\ref{eq:stationary-gpe}) gives
\begin{equation}
    \mu
    =
    U_0+g_{\rm eff}n_{\rm ref},
    \qquad
    \mu_{\rm loc}
    =
    g_{\rm eff}n_{\rm ref},
\end{equation}
where $\mu_{\rm loc}=\mu-U_0$ is the local interaction-energy scale and is positive for the repulsive interactions considered here wherever the reference density is nonzero.

Now consider a small spatial variation around this reference state,
\begin{equation}
    \psi(\mathbf r)
    =
    \psi_{\rm ref}
    +
    \delta\psi(\mathbf r)
    =
    \sqrt{n_{\rm ref}}
    +
    \delta\psi(\mathbf r).
\end{equation}
The variation can be decomposed into plane waves,
\begin{equation}
    \delta\psi(\mathbf r)
    =
    \frac{1}{(2\pi)^3}
    \int
    \widetilde{\delta\psi}(\mathbf k)
    e^{i\mathbf k\cdot\mathbf r}
    \,d^3\mathbf k.
\end{equation}
Each plane wave satisfies
\begin{equation}
    -\frac{\hbar^2}{2m}
    \nabla^2e^{i\mathbf k\cdot\mathbf r}
    =
    \frac{\hbar^2|\mathbf k|^2}{2m}
    e^{i\mathbf k\cdot\mathbf r}.
\end{equation}

Let $L_{\mathbf k}=1/|\mathbf k|$ be the distance along $\mathbf k$ over which the plane-wave phase changes by one radian, which is the wavelength divided by $2\pi$. The kinetic-energy scale of the variation is then
\begin{equation}
    E_{\rm kin}(L_{\mathbf k})
    =
    \frac{\hbar^2}{2mL_{\mathbf k}^2}.
\end{equation}
Shorter $L_{\mathbf k}$ gives a larger kinetic-energy scale, so the grid must resolve variations for which kinetic effects become comparable to the local interaction energy and can no longer be neglected.

\paragraph{Healing-Length Scale.}
The healing length $\xi$ is defined by equating the kinetic-energy scale to the local interaction-energy scale,
\begin{equation}
    \frac{\hbar^2}{2m\xi^2}
    =
    \mu_{\rm loc},
    \qquad
    \xi
    =
    \frac{\hbar}{\sqrt{2m\mu_{\rm loc}}}.
    \label{eq:healing-length}
\end{equation}
For $L_{\mathbf k}\gg\xi$, the kinetic-energy scale is small relative to the local interaction energy, consistent with the Thomas--Fermi approximation. Within a Thomas--Fermi cloud, $\mu_{\rm loc}=\mu_{\rm TF}-U(\mathbf r)$ makes $\xi$ smallest at the trap minimum, where the density is highest. Approaching the boundary from inside, $\mu_{\rm loc}\to0$ and $\xi$ diverges, indicating the breakdown of the approximation.

Taking the potential at the trap minimum to be zero, we define the initial central healing length as
\begin{equation}
    \xi_0
    =
    \frac{\hbar}{\sqrt{2m\mu_{\rm TF}(0)}},
\end{equation}
which provides the reference for predicting the minimum healing length during transport (Appendix~\ref{appendix:tf-scaling}).

\section{GPE Hydrodynamics}
\label{appendix:gpe-hydrodynamics}

We derive the hydrodynamic equations of the GPE in the Thomas--Fermi limit.

\paragraph{Madelung Form of the GPE.}
To describe how the condensate density evolves and flows, we write the wavefunction in Madelung form,
\begin{equation}
    \psi(\mathbf r,t)
    =
    \sqrt{n(\mathbf r,t)}\,e^{iS(\mathbf r,t)},
\end{equation}
where $n(\mathbf r,t)=|\psi(\mathbf r,t)|^2$ is the probability density and $S(\mathbf r,t)$ is the phase.

The time derivative and spatial derivatives are
\begin{equation}
    \begin{aligned}
        \frac{\partial\psi}{\partial t}
        &=
        \left(
            \frac{1}{2\sqrt n}\frac{\partial n}{\partial t}
            +
            i\sqrt n\,\frac{\partial S}{\partial t}
        \right)e^{iS},\\[1ex]
        \nabla\psi
        &=
        \left(
            \nabla\sqrt n
            +
            i\sqrt n\,\nabla S
        \right)e^{iS},\\[2ex]
        \nabla^2\psi
        &=
        \left[
            \nabla^2\sqrt n
            -
            \sqrt n\,|\nabla S|^2
            +
            i\left(
                2\nabla\sqrt n\cdot\nabla S
                +
                \sqrt n\,\nabla^2S
            \right)
        \right]e^{iS}.
    \end{aligned}
\end{equation}
Substituting these into the GPE and dividing by $\sqrt n\,e^{iS}$ gives
\begin{equation}
    \begin{aligned}
        \frac{i\hbar}{2n}\frac{\partial n}{\partial t}
        -
        \hbar\frac{\partial S}{\partial t}
        &=
        -\frac{\hbar^2}{2m}
        \left[
            \frac{\nabla^2\sqrt n}{\sqrt n}
            -
            |\nabla S|^2
            +
            i\left(
                \frac{\nabla n\cdot\nabla S}{n}
                +
                \nabla^2S
            \right)
        \right]\\[1ex]
        &\qquad
        +
        U
        +
        \,g_{\rm eff}\,n.
    \end{aligned}
    \label{eq:tf-hd-gpe}
\end{equation}
We separate the real and imaginary parts to obtain two coupled equations for the density and phase.

\paragraph{Continuity Equation.}
The imaginary part of Eq.~\eqref{eq:tf-hd-gpe} determines the density evolution,
\begin{equation}
    \begin{aligned}
        \frac{\partial n}{\partial t}
        &=
        -\frac{\hbar}{m}
        \left(
            \nabla n\cdot\nabla S
            +
            n\nabla^2S
        \right)\\[1ex]
        &=
        -\frac{\hbar}{m}\nabla\cdot(n\nabla S).
    \end{aligned}
\end{equation}
To connect this expression to the momentum operator $\hat{\mathbf p}=-i\hbar\nabla$, we calculate the mean momentum,
\begin{equation}
    \begin{aligned}
        \langle\mathbf p\rangle
        &=
        \int\psi^*(-i\hbar\nabla)\psi\,d^3\mathbf r\\[1ex]
        &=
        \int
        \left(
            n\hbar\nabla S
            -
            \frac{i\hbar}{2}\nabla n
        \right)
        d^3\mathbf r\\[1ex]
        &=
        \int n\hbar\nabla S\,d^3\mathbf r,
    \end{aligned}
\end{equation}
where $\int\nabla n\,d^3\mathbf r=0$ for a localized cloud. We define the local flow velocity by
\begin{equation}
    m\mathbf v=\hbar\nabla S,
    \qquad
    \langle\mathbf p\rangle
    =
    \int nm\mathbf v\,d^3\mathbf r.
\end{equation}
The density evolution then becomes the continuity equation for probability,
\begin{equation}
    \boxed{
        \frac{\partial n}{\partial t}
        +
        \nabla\cdot(n\mathbf v)
        =
        0
    }.
    \label{eq:tf-hd-continuity}
\end{equation}
This equation expresses local probability conservation, meaning that the density changes as probability flows into or out of a region.

\paragraph{Quantum Euler Equation.}
The real part of Eq.~\eqref{eq:tf-hd-gpe} determines the phase evolution,
\begin{equation}
    \begin{aligned}
        -\hbar\frac{\partial S}{\partial t}
        &=
        -
        \frac{\hbar^2}{2m}
        \frac{\nabla^2\sqrt n}{\sqrt n}
        +
        \frac{\hbar^2}{2m}|\nabla S|^2
        +
        U+g_{\rm eff}\,n\\[1ex]
        &=
        -
        \frac{\hbar^2}{2m}
        \frac{\nabla^2\sqrt n}{\sqrt n}
        +
        \frac{m}{2}|\mathbf v|^2
        +
        U+g_{\rm eff}\,n.
    \end{aligned}
\end{equation}
Taking its spatial gradient converts the phase evolution into an equation for the velocity.
\begin{equation}
    \hbar\nabla\frac{\partial S}{\partial t}
    =
    -\nabla\left(
        -
        \frac{\hbar^2}{2m}
        \frac{\nabla^2\sqrt n}{\sqrt n}
        +
        \frac{m}{2}|\mathbf v|^2
        +
        U+g_{\rm eff}\,n
    \right).
\end{equation}
Applying the vector calculus identity, the gradient of the squared speed satisfies
\begin{equation}
    \frac{1}{2}\nabla|\mathbf v|^2
    =
    (\mathbf v\cdot\nabla)\mathbf v
    +
    \mathbf v\times(\nabla\times\mathbf v)
    =
    (\mathbf v\cdot\nabla)\mathbf v,
\end{equation}
where the second equality follows because the curl of any gradient field is identically zero,
\begin{equation}
    \nabla\times\mathbf v
    =
    \frac{\hbar}{m}\nabla\times\nabla S
    =
    0.
\end{equation} 
Using $\hbar\nabla\frac{\partial S}{\partial t}=m\frac{\partial\mathbf v}{\partial t}$ therefore gives the quantum Euler equation for the velocity field,
\begin{equation}
    m\left(
        \frac{\partial\mathbf v}{\partial t}
        +
        (\mathbf v\cdot\nabla)\mathbf v
    \right)
    =
    -\nabla\left(
        U+g_{\rm eff}n
        -
        \frac{\hbar^2}{2m}
        \frac{\nabla^2\sqrt n}{\sqrt n}
    \right).
\end{equation}
This is a hydrodynamic form of the GPE, describing the condensate through density and velocity fields. 

The left-hand side of the quantum Euler equation gives the mass times the acceleration of a fluid element moving along the trajectory $\mathbf r(t)$. Applying the chain rule to the derivative of the fluid element's velocity $\mathbf v(\mathbf r(t),t)=d\mathbf r/dt$ gives
\begin{equation}
    \begin{aligned}
        \frac{d\mathbf v(\mathbf r(t),t)}{dt}
        &=
        \frac{\partial\mathbf v}{\partial t}
        +
        \frac{dx}{dt}\frac{\partial\mathbf v}{\partial x}
        +
        \frac{dy}{dt}\frac{\partial\mathbf v}{\partial y}
        +
        \frac{dz}{dt}\frac{\partial\mathbf v}{\partial z}\\
        &=
        \frac{\partial\mathbf v}{\partial t}
        +
        (\mathbf v\cdot\nabla)\mathbf v.
    \end{aligned}
    \label{eq:fluid-element-acceleration}
\end{equation}
The right-hand side of the quantum Euler equation gives the force per particle due to the external potential, interaction pressure, and quantum pressure. The quantum-pressure term arises from spatial variations in the wavefunction amplitude.

\paragraph{Thomas--Fermi Hydrodynamic Approximation.}
We adopt the Thomas--Fermi approximation, assuming that the kinetic contribution from amplitude gradients is negligible over most of the cloud. Dropping the quantum-pressure term from the quantum Euler equation gives
\begin{equation}
    \boxed{
        m\left(
            \frac{\partial\mathbf v}{\partial t}
            +
            (\mathbf v\cdot\nabla)\mathbf v
        \right)
        =
        -\nabla\left(U+g_{\rm eff}n\right)
    },
    \label{eq:tf-hd-euler}
\end{equation}
which describes flow acceleration driven by the external potential and interaction pressure. The continuity equation remains unchanged. These two equations describe the condensate dynamics in the Thomas--Fermi hydrodynamic limit.

\section{Thomas--Fermi (Ermakov) Scaling Model}
\label{appendix:tf-scaling}

We use the continuity and Euler equations (Appendix~\ref{appendix:gpe-hydrodynamics}) to derive the self-similar scaling dynamics.

\paragraph{Self-Similar Scaling.}
We seek a self-similar evolution of the initial parabolic Thomas--Fermi density profile (Eq.~\ref{eq:tf-harmonic-density}) that satisfies the hydrodynamic equations. A time-dependent matrix $B(t)$ maps initial displacements from the centroid to define the fluid-element trajectory,
\begin{equation}
    \mathbf r(\mathbf r_0,t)
    =
    \bar{\mathbf r}(t)
    +
    B(t)[\mathbf r_0-\bar{\mathbf r}(0)],
    \qquad
    B(0)=\mathbb I.
    \label{eq:self-similar-fluid-trajectory}
\end{equation}
Differentiating this trajectory and eliminating $\mathbf r_0$ using $\mathbf r_0-\bar{\mathbf r}(0)=B(t)^{-1}[\mathbf r-\bar{\mathbf r}(t)]$ gives the velocity field
\begin{equation}
    \mathbf v(\mathbf r,t)
    =
    \dot{\bar{\mathbf r}}(t)
    +
    \dot B(t)B(t)^{-1}
    [\mathbf r-\bar{\mathbf r}(t)].
    \label{eq:self-similar-velocity-field}
\end{equation}
At $t=0$, the cloud is stationary at the trap minimum, giving
\begin{equation}
    \dot B(0)=0,
    \qquad
    \bar{\mathbf r}(0)=\mathbf r_{\min}(0),
    \qquad
    \dot{\bar{\mathbf r}}(0)=0.
\end{equation}
Along the trajectory $\mathbf r(\mathbf r_0,t)$, the density evolves according to
\begin{equation}
    \frac{d}{dt}n(\mathbf r(\mathbf r_0,t),t)
    =
    \frac{\partial n}{\partial t}
    +
    \frac{\partial n}{\partial x}\frac{dx}{dt}
    +
    \frac{\partial n}{\partial y}\frac{dy}{dt}
    +
    \frac{\partial n}{\partial z}\frac{dz}{dt}
    =
    \frac{\partial n}{\partial t}
    +
    \mathbf v\cdot\nabla n.
\end{equation}
From here, $n$ denotes the density evaluated along this trajectory, and $d/dt$ follows the fluid element. Expanding $\nabla\cdot(n\mathbf v)$ in the continuity equation (Eq.~\ref{eq:tf-hd-continuity}) gives
\begin{equation}
    \frac{\partial n}{\partial t}
    +
    \nabla \cdot (n\mathbf{v})
    =
    \frac{\partial n}{\partial t}
    +
    \mathbf v\cdot\nabla n
    +
    n\nabla\cdot\mathbf v
    =
    0.
\end{equation}
Hence, the density evolution depends on the divergence of the velocity field $\nabla\cdot\mathbf v$,
\begin{equation}
    \frac{dn}{dt}
    =
    -n\nabla\cdot\mathbf v.
\end{equation}
Suppressing the time argument of $B(t)$ for brevity, we take the divergence of Eq.~\ref{eq:self-similar-velocity-field},
\begin{equation}
    \nabla\cdot\mathbf v
    =
    \nabla\cdot\left(
        \dot{\bar{\mathbf r}}(t)
        +
        \dot BB^{-1}
        [\mathbf r-\bar{\mathbf r}(t)]
    \right)
    =
    \nabla\cdot\left(
        \dot BB^{-1}\mathbf r
    \right),
\end{equation}
where $B(t)$, the centroid $\bar{\mathbf r}(t)$, and its velocity $\dot{\bar{\mathbf r}}(t)$ depend only on time, so the terms independent of $\mathbf r$ have zero spatial divergence. At fixed $t$, write $A=\dot BB^{-1}$. Then,
\begin{equation}
    \begin{aligned}
        \nabla\cdot\mathbf v
        &=
        \nabla\cdot\left(
            \dot BB^{-1}\mathbf r
        \right)
        =
        \nabla\cdot\left(
            A \mathbf r
        \right)
        =
        \nabla\cdot
        \begin{pmatrix}
            A_{11}x + A_{12}y + A_{13}z \\
            A_{21}x + A_{22}y + A_{23}z \\
            A_{31}x + A_{32}y + A_{33}z
        \end{pmatrix} \\
        &=\frac{\partial}{\partial x}\left(A_{11}x + A_{12}y + A_{13}z\right)
        + \frac{\partial}{\partial y}\left(A_{21}x + A_{22}y + A_{23}z\right) 
        + \frac{\partial}{\partial z}\left(A_{31}x + A_{32}y + A_{33}z\right) \\
        &=
        A_{11} + A_{22} + A_{33}
        =
        \operatorname{tr}\!\left(A\right)
        =
        \operatorname{tr}\!\left(\dot BB^{-1}\right)
    \end{aligned}
\end{equation} 
Jacobi’s formula gives the derivative of a matrix’s determinant. For an invertible matrix $B$,
\begin{equation}
    \frac{d}{dt}\det B
    =
    \det B\,
    \operatorname{tr}\!\left(\dot BB^{-1}\right).
\end{equation}
The divergence becomes
\begin{equation}
    \nabla\cdot\mathbf v
    =
    \operatorname{tr}\!\left(\dot B B^{-1}\right)
    =
    \frac{1}{\det B} \frac{d}{dt} \det B.
\end{equation}
Combining these results gives
\begin{equation}
    \frac{d}{dt}
    \left[
        n(\mathbf r(\mathbf r_0,t),t)\det B(t)
    \right]
    =
    \det B(t)
    \left[
        \frac{dn}{dt}
        +
        n\nabla\cdot\mathbf v
    \right]
    =
    0.
\end{equation}
Thus, $n(\mathbf r(\mathbf r_0, t),t)\det B(t)$ is constant along each trajectory. At $t=0$, this constant is $n(\mathbf r_0,0)$. Hence,
\begin{equation}
    \boxed{
        n(\mathbf r(\mathbf r_0, t),t)
        =
        \frac{1}{\det B(t)}
        n(\mathbf r_0,0)
    }.
    \label{eq:self-similar-continuity}
\end{equation}
Throughout the cloud, the density scales inversely with the volume factor $\det B(t)$.

\paragraph{Matrix Ermakov Equation.}
We derive the evolution of $B(t)$. The acceleration relation (Eq.~\ref{eq:fluid-element-acceleration}) gives
\begin{equation}
    \ddot{\mathbf r}(\mathbf r_0,t)
    =
    \frac{d}{dt}
    \mathbf v(\mathbf r(\mathbf r_0,t),t)
    =
    \left.
    \left[
        \frac{\partial\mathbf v}{\partial t}
        +
        (\mathbf v\cdot\nabla)\mathbf v
    \right]
    \right|_{\mathbf r=\mathbf r(\mathbf r_0,t)}.
\end{equation}
The vertical bar denotes evaluation at $\mathbf r(\mathbf r_0,t)$, where the Euler equation (Eq.~\ref{eq:tf-hd-euler}) becomes
\begin{equation}
    m\ddot{\mathbf r}(\mathbf r_0,t)
    =
    -\left.
    \nabla U
    \right|_{\mathbf r=\mathbf r(\mathbf r_0,t)}
    -
    \left.
    \nabla(g_{\rm eff}n)
    \right|_{\mathbf r=\mathbf r(\mathbf r_0,t)}.
\end{equation}
The harmonic trapping force follows directly from the trajectory (Eq.~\ref{eq:self-similar-fluid-trajectory}),
\begin{equation}
    \begin{aligned}
        -\left.
        \nabla U
        \right|_{\mathbf r=\mathbf r(\mathbf r_0,t)}
        &=
        -H(t)
        [\mathbf r(\mathbf r_0,t)-\mathbf r_{\min}(t)]\\
        &=
        -H(t)
        [\bar{\mathbf r}(t)-\mathbf r_{\min}(t)]
        -
        H(t)B(t)
        [\mathbf r_0-\bar{\mathbf r}(0)].
    \end{aligned}
\end{equation}
To evaluate the interaction-pressure force, we first use the initial equilibrium condition. At $t=0$, the fluid elements have zero acceleration, so the interaction-pressure force balances the trapping force,
\begin{equation}
    -\nabla_{\mathbf r_0}
    [g_{\rm eff}n(\mathbf r_0,0)]
    =
    \nabla_{\mathbf r_0}U(\mathbf r_0,0)
    =
    H(0)
    [\mathbf r_0-\bar{\mathbf r}(0)],
    \label{eq:initial-force-balance}
\end{equation}
where $\nabla_{\mathbf r_0}$ differentiates with respect to the initial position, and $\bar{\mathbf r}(0)=\mathbf r_{\min}(0)$ and $B(0)=\mathbb I$. We transform this initial density gradient to the current coordinates. The inverse trajectory relation is
\begin{equation}
    \mathbf r_0(\mathbf r,t)
    =
    \bar{\mathbf r}(0)
    +
    B(t)^{-1}[\mathbf r-\bar{\mathbf r}(t)],
    \qquad
    \frac{\partial\mathbf r_0}{\partial\mathbf r}
    =
    B(t)^{-1},
\end{equation}
where $\mathbf r_0(\mathbf r,t)$ identifies the initial position of the fluid element currently at $\mathbf r$. Since $\det B(t)$ is spatially uniform, differentiating the self-similar density relation (Eq.~\ref{eq:self-similar-continuity}) gives, for each Cartesian coordinate,
\begin{equation}
    \begin{aligned}
        \left.
        \frac{\partial}{\partial r_i}
        [g_{\rm eff}n(\mathbf r,t)]
        \right|_{\mathbf r=\mathbf r(\mathbf r_0,t)}
        &=
        \frac{1}{\det B(t)}
        \sum_j
        \frac{\partial[g_{\rm eff}n(\mathbf r_0,0)]}
            {\partial(r_0)_j}
        \frac{\partial(r_0)_j}{\partial r_i}\\
        \llap{In vector form:}\qquad
        \left.
        \nabla[g_{\rm eff}n(\mathbf r,t)]
        \right|_{\mathbf r=\mathbf r(\mathbf r_0,t)}
        &=
        \frac{1}{\det B(t)}
        \left(
            \frac{\partial\mathbf r_0}{\partial\mathbf r}
        \right)^{\!\top}
        \nabla_{\mathbf r_0}
        [g_{\rm eff}n(\mathbf r_0,0)].
    \end{aligned}
\end{equation}
Substituting the initial force balance (Eq.~\ref{eq:initial-force-balance}) gives the interaction-pressure force during transport,
\begin{equation}
    -\left.
    \nabla[g_{\rm eff}n(\mathbf r,t)]
    \right|_{\mathbf r=\mathbf r(\mathbf r_0,t)}
    =
    \frac{1}{\det B(t)}
    B(t)^{-\top}
    H(0)
    [\mathbf r_0-\bar{\mathbf r}(0)].
\end{equation}
Differentiating the fluid-element trajectory twice gives $\ddot{\mathbf r}(\mathbf r_0,t) = \ddot{\bar{\mathbf r}}(t) + \ddot B(t) [\mathbf r_0-\bar{\mathbf r}(0)]$. Combining this acceleration and both forces in the Euler equation, with $K(t)=H(t)/m$, gives
\begin{equation}
    \ddot{\bar{\mathbf r}}(t)
    +
    \ddot B(t)[\mathbf r_0-\bar{\mathbf r}(0)]
    =
    -K(t)[\bar{\mathbf r}(t)-\mathbf r_{\min}(t)]
    +
    \left[
        -K(t)B(t)
        +
        \frac{1}{\det B(t)}
        B(t)^{-\top}K(0)
    \right]
    [\mathbf r_0-\bar{\mathbf r}(0)].
\end{equation}
Because this holds for every initial position $\mathbf r_0$, the spatially uniform terms and the coefficients of $[\mathbf r_0-\bar{\mathbf r}(0)]$ must agree separately. The centroid and scaling matrix evolve independently according to
\begin{equation}
    \begin{aligned}
        \ddot{\bar{\mathbf r}}(t)
        &=
        -K(t)
        [\bar{\mathbf r}(t)-\mathbf r_{\min}(t)],\\
        \ddot B(t)
        &=
        -K(t)B(t)
        +
        \frac{1}{\det B(t)}
        B(t)^{-\top}K(0).
    \end{aligned}
\end{equation}
The second equation is the matrix Ermakov equation.

\paragraph{Laboratory-Axis Extents.}
Using the scaled coordinates $\mathbf u$ (Eq.~\ref{eq:tf-scaled-coords}) at $t=0$, points in the initial Thomas--Fermi ellipsoid are parameterized as
\begin{equation}
    \mathbf r_0-\bar{\mathbf r}(0)
    =
    Q(0)
    \operatorname{diag}\!\left(
        R_{{\rm TF},1}(0),
        R_{{\rm TF},2}(0),
        R_{{\rm TF},3}(0)
    \right)
    \mathbf u,
    \qquad
    \mathbf u^\top\mathbf u\leq1,
\end{equation}
where the columns of $Q(0)=[\hat{\mathbf e}_1(0)\ \hat{\mathbf e}_2(0)\ \hat{\mathbf e}_3(0)]$ are the initial principal-axis unit vectors expressed in laboratory coordinates. The diagonal matrix scales these vectors by the TF radii, and multiplication by $\mathbf u$ gives the initial displacement in laboratory coordinates.

The scaling transformation gives
\begin{equation}
    \begin{aligned}
        \mathbf r(\mathbf r_0,t)-\bar{\mathbf r}(t)
        &=
        B(t)[\mathbf r_0-\bar{\mathbf r}(0)]
        =
        \mathcal R(t)\mathbf u,\\
        \mathcal R(t)
        &=
        B(t)Q(0)
        \operatorname{diag}\!\left(
            R_{{\rm TF},1}(0),
            R_{{\rm TF},2}(0),
            R_{{\rm TF},3}(0)
        \right).
    \end{aligned}
\end{equation}
Each row of $\mathcal R(t)$ determines the displacement along one laboratory axis. Its Euclidean norm gives the largest displacement magnitude for $|\mathbf u|\leq1$, so the maximum half-extent during transport is
\begin{equation}
    E_i^{\rm Erm}
    =
    \max_{0\leq t\leq T}
    \sqrt{
        \sum_{k=1}^{3}
        \mathcal R_{ik}^2(t)
    }.
\end{equation}
These extents give the Ermakov term in Eq.~\ref{eq:box-sizing}. For a fixed trap schedule, the atom-number dependence of $\mathcal R(t)$ comes from the initial TF radii. The equilibrium relations (Appendix~\ref{appendix:tf-equilibrium}) give $\mu_{\rm TF}\propto g_{\rm eff}^{2/5}$ and $R_{{\rm TF},k}\propto\sqrt{\mu_{\rm TF}}$. With $g_{\rm eff}\simeq Ng_0$, the radii and hence $E_i^{\rm Erm}$ scale as $N^{1/5}$.

\paragraph{Healing Length During Transport.}
Self-similar scaling (Eq.~\ref{eq:self-similar-continuity}) changes the density by $1/\det B(t)$. Since $\xi\propto n^{-1/2}$ (Appendix~\ref{appendix:healing-length}), the predicted healing length evolves as
\begin{equation}
    \xi(t)
    =
    \xi_0\sqrt{\det B(t)}.
\end{equation}
Since $\xi_0$ is the initial central healing length, $\xi(t)$ gives the central value, which remains the smallest within the self-similarly evolving cloud. The minimum predicted value during the transport is therefore
\begin{equation}
    \xi_{\min}
    =
    \min_{0\leq t\leq T}\xi(t)
    =
    \xi_0
    \sqrt{
        \min_{0\leq t\leq T}\det B(t)
    }.
\end{equation}
At each time, $\xi(t)$ is the distance at which the kinetic contribution from an amplitude variation becomes comparable to the local interaction energy. Thus, $\xi_{\min}$ is the smallest predicted length scale at which kinetic and interaction energies are comparable during transport. To resolve this scale throughout transport, we choose each grid interval to be no larger than $\xi_{\min}/2$ (Eq.~\ref{eq:spatial-resolution}).

\clearpage
\printbibliography

\end{document}